\documentclass[conference]{IEEEtran}
\IEEEoverridecommandlockouts
\usepackage{cite}
\usepackage{amsmath,amssymb,amsfonts}
\usepackage{algorithmic}
\usepackage{graphicx}
\usepackage{textcomp}
\usepackage{xcolor}
\def\BibTeX{{\rm B\kern-.05em{\sc i\kern-.025em b}\kern-.08emT\kern-.1667em\lower.7ex\hbox{E}\kern-.125emX}}

\usepackage{csquotes}

\usepackage{caption}
\usepackage{subcaption}

\usepackage{graphicx}
\usepackage{listings}
\usepackage{tikz}
\usepackage[english]{babel}
\usepackage{amsthm}

\usepackage{hyperref}

\usepackage{amsfonts}
\usepackage{xcolor}
\usepackage{breqn}
\usepackage{amsmath}

\newcommand{\meritrank}{{\textsc{MeritRank}}}

\begin{document}

\title{MeritRank: Sybil Tolerant Reputation for Merit-based Tokenomics*\thanks{*pre-print BRAINS conference, Paris, September 27-30, 2022}}

\author{
 \IEEEauthorblockN{Bulat Nasrulin, Georgy Ishmaev, Johan Pouwelse}
 \IEEEauthorblockA{Delft University of Technology
 \\\{b.nasrulin, g.ishmaev, j.a.pouwelse,\}@tudelft.nl}
}

\maketitle

\theoremstyle{definition}
\newtheorem{definition}{Definition}[section]
\newtheorem{theorem}{Theorem}[section]

\newcounter{subdef}[definition]
\renewcommand{\thesubdef}{\thedefinition\alph{subdef}}

\newenvironment{superdefinition}
  {\refstepcounter{definition}}
  {}

\newenvironment{subdefinition}[1]
  {\refstepcounter{subdef}\noindent\textbf{Definition \thesubdef} \textbf{(#1)}\itshape}
  {}

\begin{abstract}
Decentralized reputation systems are emerging as promising mechanisms to enhance the effectiveness of token-based economies. Unlike traditional monetary incentives, these systems reward participants based on the actual value of their contributions to the network. However, the advantages and challenges associated with such systems remain largely unexplored. In this work, we investigate the inherent trade-offs in designing a decentralized reputation system that is simultaneously generalizable, trustless, and Sybil-resistant. Specifically, “generalizable” means that the system can assess various types of contributions across different contexts, “trustless” indicates that it functions without the need for a central authority to oversee reputations, and “Sybil-resistant” refers to its ability to withstand manipulations by fake identities, i.e., Sybil attacks.

We propose \meritrank{}, a Sybil-tolerant reputation system based on feedback aggregation from participants. Instead of entirely preventing Sybil attacks, our approach effectively limits the benefits that attackers can gain from such strategies. 	This is achieved by reducing the perceived value of the attacker’s and Sybil nodes’ contributions through the application of decay mechanisms—specifically, transitivity decay, connectivity decay, and epoch decay. Using a dataset of participant interactions in MakerDAO, we conducted experiments to demonstrate the Sybil tolerance of \meritrank{}.
\end{abstract}

\section{Introduction}

Reputation mechanisms in blockchain applications provide numerous desirable properties as system components. These mechanisms can be employed at various layers of blockchain systems.  At the infrastructure level, anyone can act as a relay node, competing based on reputation for how quickly and neutrally they distribute transactions~\cite{noauthor_mev-boost_2021}.  At the protocol layer, Delegated Proof-of-Stake (DPoS)~\cite{nguyen_proof--stake_2019} utilizes mechanisms to detect and punish undesirable behaviors, such as double voting on blocks~\cite{li2017securing}. At the application layer, Decentralized Autonomous Organizations (DAOs)~\cite{aragon_dao_reputation_2022,noauthor_coordinape_nodate,noauthor_sourcecred_nodate,rea_colony_2020} employ rewards for participants based on their contributions.

Token-based incentives have emerged as prominent mechanisms in blockchain protocols, addressing several incentivization issues in peer-to-peer networks, such as ensuring network liveness, enhancing network security, and supporting open-source software maintenance. This exploration of incentives has given rise to a new subfield within blockchain design, often referred to as \emph{tokenomics}. Consequently, these mechanisms are integral to most significant blockchain applications, utilized at various abstraction levels~\cite{freni_2020}.

However, the limitations of traditional token-based incentives are increasingly evident through empirical evidence. First, misalignment of incentives often occurs, where monetary rewards can lead to conflicts among participants within complex systems~\cite{garg_makerdao_2019,liam_j_how_2022}. These conflicts arise because such rewards do not always reflect the actual value of contributions, fostering self-interested behaviors that can undermine collective goals. Second, governance models relying solely on monetary incentives lack the robustness necessary for effective decentralized decision-making~\cite{daian_-chain_2018}. Lastly, these incentives tend to disproportionately benefit larger stakeholders, risking the re-centralization of decentralized networks~\cite{martinazzi_evolving_2020}.

An alternative approach is to reward participants based on their contributions or \emph{merits} through a reputation system~\cite{nowak_five_2006}. Examples of contributions include computational work, proof of bandwidth, proof of storage, participation in DAO governance forums, open-source code development, and discussions in chat messengers. DAOs often establish treasuries to incentivize contributions by periodically rewarding the most active participants with tokens. However, this approach faces significant challenges due to Sybil attacks~\cite{douceur2002sybil, dinger_defending_2006}, where attackers use multiple fake identities to manipulate reputations and deplete the treasury. Such attacks are a major obstacle to deploying advanced reputation systems intended to surpass the simplistic economic models often found in current tokenomics discussions~\cite{de_filippi_reputation_2021, weyl_2022}.


While some previous work has proposed Sybil-resistant reputation algorithms~\cite{meulpolder2009bartercast, otte2020trustchain, liu2016personalized}, these solutions only partially mitigate Sybil attacks and remain vulnerable to reputation manipulation. First, Sybil identities are often indistinguishable from legitimate identities~\cite{alvisi2013sok}, making many Sybil-detection algorithms ineffective. Second, achieving full Sybil resistance is challenging because Sybil identities can still receive positive feedback from legitimate participants. This positive feedback allows Sybil identities to blend in and accumulate reputation, thereby making Sybil attacks beneficial. Such attacks are prevalent in social networks where interactions and feedback can be easily manipulated~\cite{viswanath2010analysis}. 

We observe that Sybil attacks become advantageous when the attacker makes only minimal contributions, yet both the attacker and their Sybil identities receive positive reputations and token rewards. This dynamic can undermine the integrity of reputation systems. 

Rather than trying to fully prevent Sybil attacks or detect Sybil identities, our approach focuses on limiting the benefits that attackers can gain from such attacks, thereby making them less attractive and practical. To achieve this, we propose \meritrank{}, a Sybil-tolerant, aggregated feedback reputations compatible with token-based incentive systems. \meritrank{} achieves Sybil-Tolerance by bounding the benefit of the attack through the use of personalized reputation and decay heuristics to reduce the perceived value of attacker's contributions. 

In MeritRank, we introduce and test three types of decay mechanisms, building upon findings from previous research~\cite{sybil_proofness_2021, liu2016personalized}, to enhance the resilience of reputation systems against Sybil attacks. \textit{A seed node} is a trusted starting point from which reputations are calculated, serving as a reference for evaluating other nodes’ contributions. The three decays are \textit{transitivity decay}, \textit{connectivity decay}, and \textit{epoch decay}. Transitivity decay reduces the perceived value of contributions based on their distance from the seed node. Connectivity decay lowers the value of contributions from nodes that rely on a single path or a limited set of connections to the seed node, such as those connected only through bridges. Epoch decay decreases the value of older contributions. These mechanisms reduce the effectiveness of Sybil attacks, making them less attractive. However, this comes at a cost: honest participants might receive lower reputations due to the decay of their perceived contributions. Despite this drawback, we will see that the improved resilience of the system justifies the trade-off.

This chapter makes both theoretical and practical contributions, which are as follows:
\begin{itemize}
\item We analyze and formulate the general trade-offs between desirable properties of reputation in decentralized settings by presenting a \textit{decentralized reputation trilemma} in Section \ref{sec:background}.
\item We present the formalization of \meritrank{}, along with three types of decay mechanisms—transitivity decay, connectivity decay, and epoch decay—designed to achieve tolerance against Sybil attacks in Section \ref{sec:meritrank}. We evaluate \meritrank{} using a dataset spanning over 150 weeks of user interactions within the leading decentralized autonomous organization, MakerDAO, as discussed in Section~\ref{sec:experiments}. Our experiments demonstrate that transitivity decay and connectivity decay significantly enhance the Sybil tolerance of reputation algorithms. However, we observe that the popular heuristic, epoch decay, used in other reputation mechanisms~\cite{rea_colony_2020}, does not improve Sybil tolerance.
\end{itemize}

\section{Background and Related Work} \label{sec:background}

Reputation systems in decentralized environments have been proposed for various applications, and as general models in peer-to-peer systems~\cite{walsh_experience_2006,delaviz_sybilres_2012,hendrikx_reputation_2015,bellini_blockchain-based_2020,gurtler_sok_2021, sybil_proofness_2021}. Accordingly, the limitations of these solutions are relatively well-understood. Some of these limitations include scalability~\cite{bellini_blockchain-based_2020}, contextual accuracy~\cite{hendrikx_reputation_2015}, reliance on partially trusted setups~\cite{hendrikx_reputation_2015,bellini_blockchain-based_2020}, vulnerability to misreporting attacks~\cite{cheng2005sybilproof,hoffman2009survey,viswanath2012exploring,koutrouli_taxonomy_2012,seuken2014work}, and  privacy trade-offs~\cite{gurtler_sok_2021}. 

However, there is an identifiable research gap regarding the general trade-offs inherent to any reputation system implemented in decentralized environments. We are able to identify only a small number of surveys on reputation solutions in decentralized settings~\cite{hendrikx_reputation_2015,bellini_blockchain-based_2020}, with limited comparative trade-off analysis~\cite{gurtler_sok_2021}.  The absence of engineering research on reputation in the context of tokenomics and DAOs can be attributed to the novelty of these problems~\cite{el_faqir_overview_2020}, which have only recently gained attention in the academic and engineering communities.

\subsection{Decentralized Reputation Trilemma}

\begin{figure}[t]
\centering
\includegraphics[width=0.7\linewidth]{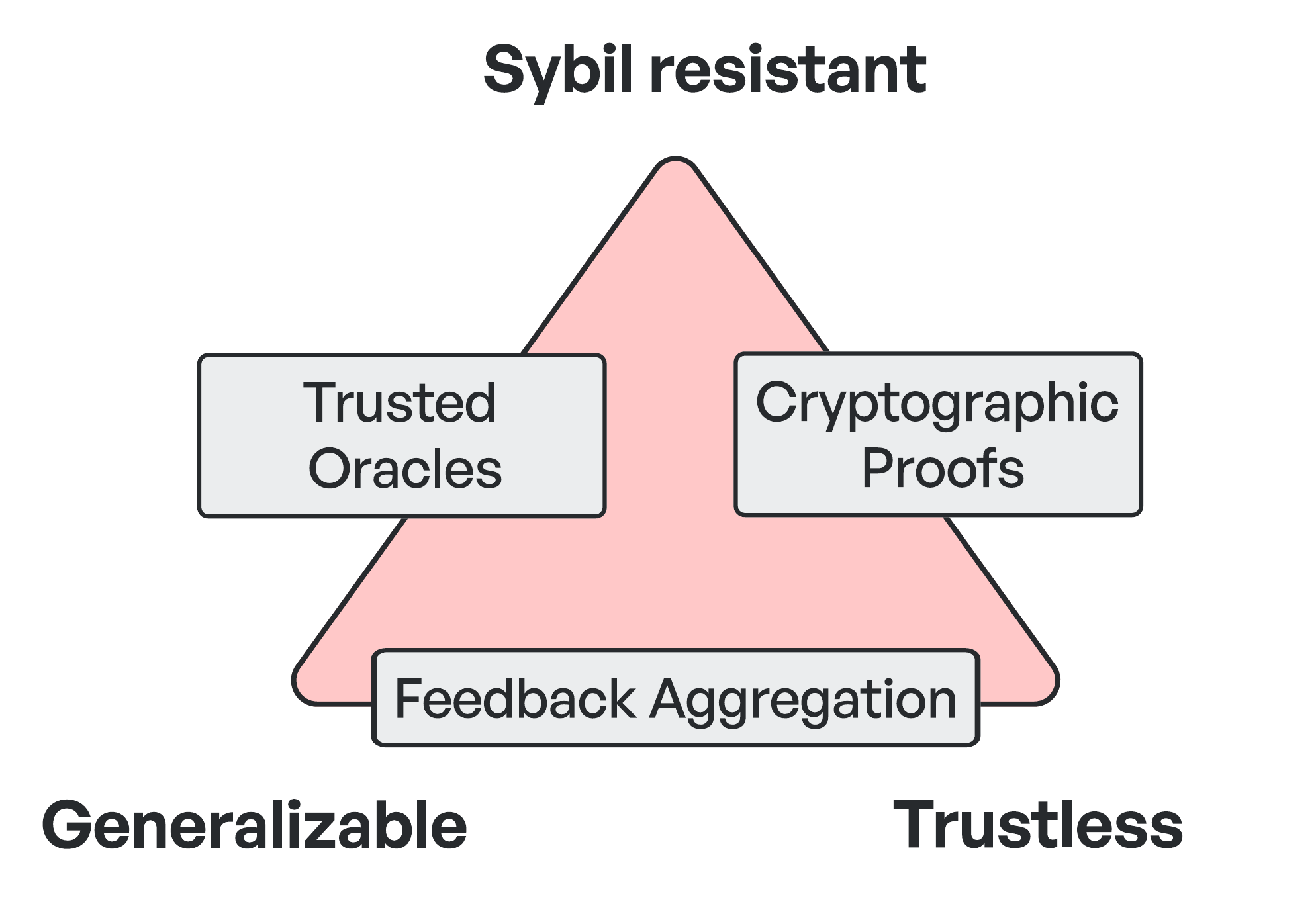}
\caption{The “Decentralized Reputation Trilemma” illustrates the irreconcilability of three desirable properties: Generalizable, Sybil-resistant, and Trustless. The triangle’s edges represent approaches that sacrifice one property: Trusted Oracles sacrifice Trustlessness, Cryptographic Proofs sacrifice Generalizability, and Feedback Aggregation sacrifices Sybil resistance.}
\label{fig:reputation_triangle}
\end{figure}

We formulate the inherent trade-offs in decentralized reputation systems as a conjecture on the irreconcilability of three desirable properties. This trilemma is depicted in Figure \ref{fig:reputation_triangle}.  A reputation system cannot simultaneously embody all three attributes: \textit{Generalizability}, \textit{Sybil resistance}, and \textit{Trustlessness}.

\textit{Generalizability} refers to the system’s ability to evaluate and fairly assess a wide variety of contributions, including both technical and human-centric activities. This flexibility allows the system to adapt to different types of participant roles and behaviors, without being limited to specific, easily measurable tasks. A generalizable system must also scale efficiently, maintaining performance as the network grows, while providing accurate, context-aware reputation scores that account for the nuances of each contribution.

A fully \textit{Sybil-resistant} system prevents attackers from manipulating reputation metrics through the creation and control of multiple fake identities~\cite{douceur2002sybil}. This is crucial for maintaining the integrity of the reputation system, as it ensures that reputation scores genuinely reflect legitimate contributions.

The \textit{Trustless} property means that the system’s reputation accounting and evaluation processes do not rely on any single trusted entity. Instead, it leverages decentralized protocols and mechanisms to ensure that all operations are transparent, verifiable, and resistant to tampering or biased influence. This decentralization enhances the system’s security and fairness by minimizing the risk of corruption or failure associated with relying on trusted intermediaries.

These three properties form a trilemma because optimizing for two typically compromises the third. To better understand how different approaches navigate these trade-offs, we examine three primary methods used in decentralized reputation systems: Trusted Oracles, Cryptographic Proofs, and Feedback Aggregation.

\textbf{Trusted Oracles}. This method employs trusted oracles to monitor and calculate reputation scores, relying on predefined reputation functions and participant actions as inputs. Such a system is inherently generalizable, as oracles can process various inputs to determine reputation scores~\cite{bellini_blockchain-based_2020}. The challenge of Sybil resistance is managed by entrusting oracles with the task of detecting and verifying the identities of participants, thereby preventing Sybil attacks~\cite{siddarth_who_2020}. However, this reliance on oracles compromises the trustless nature of the system. Participants must trust the oracles to accurately process inputs and honestly calculate reputation scores. Moreover, the dependence on oracles introduces potential points of failure that adversaries could exploit, as observed with price oracles in decentralized finance systems~\cite{eskandari_sok_2021}.

Efforts to incorporate trustless characteristics within oracles have led to the creation of peer prediction markets. In these systems, participants estimate and report outcomes, with their rewards determined by how well their reports align with those of others. This method is designed to be self-regulating and decentralized, reducing reliance on a single authoritative oracle. However, in scenarios where quick or highly specialized assessments are needed, the peer prediction method might struggle to provide accurate and timely evaluations, thus limiting its generalizability~\cite{cai_truth-inducing_2020}.

\textbf{Cryptographic Proofs}. With this approach, peers document their own and others’ reputation scores by generating cryptographic proofs of contributions, which are then disseminated across the network. Upon validation of these proofs, participants adjust their reputation scores accordingly. The requirement for verifiable proof of contribution inherently reduces the feasibility of Sybil attacks, as each entity must substantiate its contributions through cryptographic evidence. This method also achieves trustlessness, as it eliminates the need for a singular authoritative body overseeing reputation accounting.

However, the applicability of cryptographic proofs to reputation systems faces significant constraints, particularly regarding generalizability. While certain contributions—such as computational work, proof of bandwidth, or proof of storage—are well-suited for cryptographic validation, this framework struggles to accommodate a broader spectrum of collaborative efforts and human-centric contributions. Many types of cooperative work cannot be easily verified cryptographically, limiting the range of activities that can be effectively accounted for in such a system. Moreover, even when cryptographic proofing is conceptually applicable, its practical implementation faces challenges related to scalability. The extensive overhead required for generating, distributing, and validating cryptographic proofs can significantly strain system resources, impacting both scalability and efficiency~\cite{wust2018you}.

\textbf{Feedback Aggregation} calculates an individual’s reputation by collecting feedback directly from other participants on the perceived value of that individual’s contributions. The system then aggregates this feedback to determine the overall reputation score. This method is generalizable and trustless. By facilitating peer-to-peer feedback, the system can adapt to diverse application-specific scenarios, allowing participants to provide nuanced, context-specific feedback on a wide array of interactions or contributions. This adaptability makes it exceptionally suitable for environments where direct peer evaluation is possible and relevant. Additionally, unlike approaches reliant on cryptographic proofs, feedback aggregation does not incur significant overhead costs, making it a more resource-efficient option.

However, this method is susceptible to manipulation, particularly through Sybil attacks. In such attacks, a malicious entity creates multiple fake identities to flood the system with false feedback or fraudulent contributions, thereby distorting the reputation scores~\cite{douceur2002sybil}.

Previous research has proposed various approaches to Sybil-resistant reputation systems. For instance, EigenTrust~\cite{kamvar2003eigentrust} uses a global trust value computed through iterative aggregation of local trust scores but is susceptible to Sybil attacks if attackers can accumulate sufficient local trust. SybilGuard~\cite{yu2006sybilguard} and SybilLimit~\cite{yu2008sybillimit} leverage social network structures to limit the number of Sybil nodes accepted but rely on the assumption that the social graph is fast-mixing, which may not hold in all cases.

\subsection{Our Approach to Solve the Reputation Trilemma}\label{sec:reputation}

Observations based on the reputation trilemma necessitate a solution for reputation systems in decentralized environments that does not completely sacrifice one of the corners of this triangle. Because of the lack of generalizability of \textit{Cryptographic Proofs} and the trust assumptions inherent in \textit{Trusted Oracles}, the \textit{Feedback Aggregation} approach emerges as a viable direction. This is contingent on our ability to improve the resistance of reputation aggregation functions to Sybil attacks.

A common way to achieve Sybil resistance is to emulate a closed system trying to achieve \textit{Sybil prevention}. For example, by requiring participants to undergo identity verification processes, the system can limit the creation of fake identities. However, this approach often conflicts with the principles of privacy and decentralization inherent in open systems. In the context of open, permissionless systems, strict Sybil resistance is not entirely achievable~\cite{alvisi2013sok}. An alternative approach to Sybil resistance is based on \textit{Sybil detection}~\cite{viswanath2012canal} and the subsequent exclusion of Sybils from the system. The effectiveness of Sybil detection is generally constrained by several factors. In open, decentralized, and pseudonymous networks, identities are easily created, and there is no reliable way to link digital identities to unique real-world entities. This makes it inherently challenging to distinguish between legitimate users and Sybil identities. Attackers can create multiple identities that behave indistinguishably from honest nodes, making detection algorithms ineffective~\cite{alvisi2013sok}. Additionally, imposing strict verification measures conflicts with the principles of decentralization and user privacy.

\textit{Sybil tolerance}~\cite{viswanath2012exploring} focuses on minimizing the impact of Sybil identities rather than attempting to identify and remove them. By accepting that some Sybil identities may infiltrate the system, the focus shifts to reducing the damage they can cause. This can be achieved by implementing mechanisms that restrict the influence any single participant or group of participants can exert on the reputation system. Sybil tolerance is achieved by limiting the relative benefit that an attacker can gain through the use of Sybils. Specifically, a reputation system is considered Sybil-tolerant if, even as an attacker creates additional Sybil identities, the cumulative influence of these identities on the reputation scores remains limited.

One major limitation of Feedback Aggregation systems is their reliance on heuristic methods to determine the influence of participants, which can result in inaccurate reputation assessments. If the heuristics are too strict, legitimate users may be unfairly penalized, reducing the system’s fairness and overall effectiveness. Conversely, if the heuristics are too lenient, they may fail to sufficiently limit the influence of Sybil identities, allowing attackers to cause significant harm. Despite these inherent challenges and limitations, we adopt the Feedback Aggregation approach. Our goal is to strike a balance between overly strict and overly lenient heuristics, ensuring fairness for legitimate users while maintaining resilience against Sybil attacks.

\section{Merit-based Tokenomics}\label{sec:merit_model}

In this section, we present a general system model for merit-based tokenomics. This model describes a reward system for the participants of a generic DAO, where peers provide feedback to each other resulting in a reputation ranking that can be used to distribute token rewards from a DAO treasury proportionally to accrued reputation.  We acknowledge that reputation mechanisms in distributed systems face challenges, such as incompleteness of information about peer interactions and peer discovery. Therefore, our design accommodates these limitations by functioning effectively with only partial information.

\begin{figure}
	\center
	\includegraphics[width=0.85\linewidth]{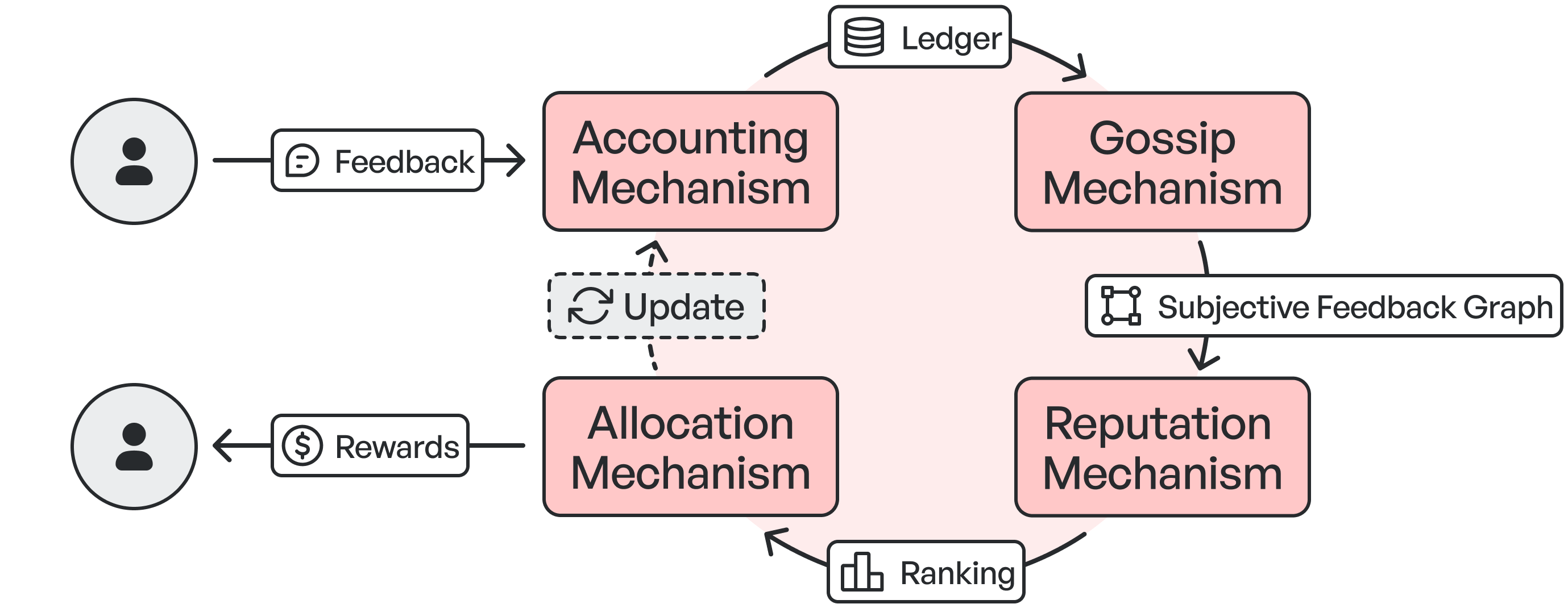}
    	\caption{Merit-Based Tokenomics system model.}
	\label{fig:model}
\end{figure}

The model is illustrated in Figure~\ref{fig:model} and includes four mechanisms. The \emph{accounting mechanism} records locally computed or recorded feedback in a personal ledger. The \emph{gossip mechanism} distributes this information to peers, resulting in the collection of indirect feedback (i.e., feedback gathered from other participants) in the form of a feedback graph. This graph is used by the \emph{reputation mechanism} to calculate participants’ reputations. Finally, the \emph{allocation mechanism} allocates rewards based on reputation rankings. This in turn might result in an update in the local ledger. The model is dynamic and operates in \emph{epochs}, with each epoch representing a discrete time step during which changes to the network structure occur, encompassing one complete iteration of the system’s feedback loop. We will now discuss each mechanism in detail. 

\subsection{Accounting Mechanism}

We assume that peers might interact through various activities, such as making direct contributions, creating content, participating in discussions, or engaging in collaborative tasks. In response to these interactions, peers provide \emph{feedback}, which can take the form of evaluations, ratings, or other reactions. This feedback is represented as a directed, weighted graph, denoted as $G = \left(V, E, w\right)$, and referred to as the \emph{feedback graph}. In this graph, each node represents a peer in the network, while each directed edge indicates feedback provided from one peer to another. The weight of an edge, defined by the function $w: V \times V \rightarrow \mathbb{R}_{\geq 0}$, assigns a non-negative value that quantifies the cumulative feedback assigned.

\begin{figure}
	\center
	\includegraphics[width=0.5\linewidth]{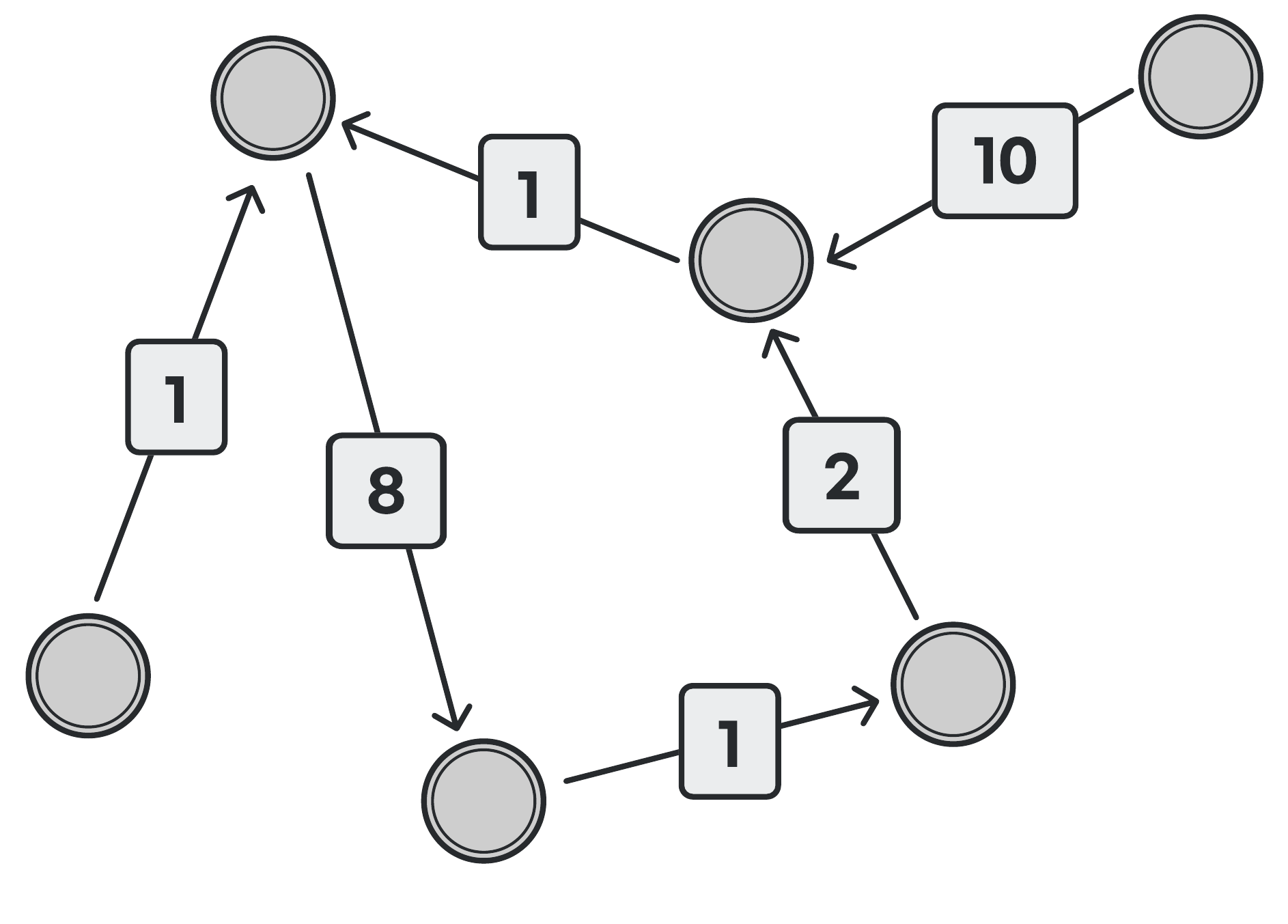}
	\caption{Example of a feedback graph. Edge weights represent the total feedback assigned by a participant about another participant.}
	\label{fig:graph}
\end{figure}

An example of a feedback graph is shown in Figure~\ref{fig:graph}, where the weights on the edges represent feedback quantified in arbitrary units. Our model is agnostic to the method of feedback generation, which can range from simple social reactions—such as likes, votes, or endorsements—to more complex algorithms that assign scores based on contributions. For instance, a “thumbs up” can be directly translated into a single unit, while systems like SourceCred~\cite{noauthor_sourcecred_nodate} compute scores for actions such as code commits or issue resolutions based on predefined metrics. Similarly, feedback can come from explicit responses (e.g., ratings or endorsements) or from implicit signals (e.g., engagement levels or contribution frequency) as seen in platforms like Discourse~\cite{discourse} and GitHub~\cite{miyazono_2020}. In both cases, the resulting value is recorded in the personal ledger of the observing peer.

\subsection{Gossip Mechanism}

Peers exchange and propagate updates to the feedback graph using a peer-to-peer gossip protocol~\cite{tarr2019secure}. Through this mechanism, feedback data recorded in each peer’s personal ledger is periodically shared across the network, allowing reputation data to evolve based on recent interactions. Each peer maintains its own \emph{subjective} view of the global feedback graph, denoted as  $G_i = (V_i, E_i, w_i)$, where  $V_i$  and  $E_i$  represent the known nodes and feedback links, and  $w_i: V_i \times V_i \rightarrow \mathbb{R}_{\geq 0}$  assigns weights based on the latest received data.

Due to network delays or incomplete data propagation, the local graph $G_i$ may differ from the views of other peers. When a peer receives new updates, such as the addition of a new edge or a change in the weight of an existing edge, it updates its local view and records this information in its personal ledger. As peers continue to share more data, these subjective graphs gradually converge, resulting in a more consistent view of the network.

\subsection{Reputation Mechanism}

We assume that peers can discover the feedback graph through mechanisms such as a gossip protocol. The reputation data  evolves with time and reflects the latest interactions by utilizing a peer-to-peer gossip protocol~\cite{tarr2019secure}. Specifically, each peer distributes the latest locally known weights in the feedback graph. Using this information, each peer $i$ constructs its \emph{subjective} feedback graph  $G_i = (V_i, E_i, w_i)$. Every time a peer receives a new edge or a weight update from the network, it records this in its personal ledger and updates its subjective feedback graph  $G_i$.

A reputation mechanism calculates and assigns reputation scores to every node in the subjective feedback graph. The reputation score reflects the level of contributions made by a node relative to others.

\begin{definition}[Reputation Score]
A reputation score  $R_i(G_i, j)$  is a non-negative value calculated and assigned to node  $j$  by the reputation mechanism of node $i$, given the subjective feedback graph  $G_i$:
\begin{equation*}
R_i(G_i, j) \in \mathbb{R}_{\geq 0} \quad \forall j \in V_i \setminus \{i\} 
\end{equation*}
\end{definition}

\subsection{Allocation Mechanism}

Reputation scores are used as inputs for an allocation mechanism that determines how rewards are distributed among a set of nodes using an allocation policy. Specifically we assume some seed node $i$, which is also part of $V$. The seed node uses its local reputation scores to distribute rewards to other nodes according to their reputations at the end of each epoch.

\begin{definition}[Allocation score]
An allocation score $A_i(G_i, j)$ is a non-negative value calculated and assigned to node $j$ by the allocation mechanism of node $i$, based on the subjective feedback graph $G_i$ and reputation scores $R_i(G_i, j)$:
\begin{equation*}
    A_i(R_i(G_i, j), j) \in \mathbb{R}_{\geq 0} \quad \forall j \in V_i\setminus{}\{i\}
\end{equation*}
\end{definition}

We do not make any specific assumptions about the allocation mechanisms but assume that nodes with higher reputation scores are more likely to receive rewards or receive larger rewards. The allocation of rewards depends on the specific seed node. If there are multiple seed nodes, a node will receive rewards separately from each seed node, and the total rewards will be the sum of these individual allocations.

One example of an allocation mechanism is \textit{winner-takes-all}, where the node with the highest reputation receives all the rewards. Another popular example is the quadratic distribution~\cite{weyl_2022}, where rewards are distributed such that each node receives a portion of the total reward pool proportional to the square root of its reputation score. This ensures that while nodes with higher reputations receive more significant rewards, the rate of increase in rewards diminishes as reputations grow, preventing disproportionate advantages.

Different allocation mechanisms will have different applications and contexts where they are most effective. However, a detailed discussion on the specific properties and implications of various allocation policies is beyond the scope of this work.

\section{Sybil Attack and Sybil Tolerance}\label{sec:sybil_attack}

In this section, we introduce the model for a Sybil attack on merit-based tokenomics, as presented in the previous section. The attacker executes the Sybil attack to inflate its reputation and subsequently gain disproportionately more rewards. We then discuss various strategies that attackers can employ to maximize the effectiveness of Sybil attacks. Finally, we present a model for Sybil tolerance, which quantifies how well a reputation mechanism can withstand a Sybil attack. We conclude by reporting on the most beneficial Sybil attack strategies against the most commonly used reputation mechanisms.

\subsection{Sybil Attack}

We model a Sybil attack as \textit{strategic}, meaning the attacker first infiltrates into the system pretending to be an honest node, receiving legitimate feedback from other honest nodes, and then executes a Sybil attack. The attacker achieves this by creating fake identities (Sybil nodes) and fake edges connecting these identities to each other (Sybil edges). The weights of the Sybil edges can be arbitrary. We refer to the created subgraph consisting of Sybil nodes and edges as \textit{Sybil region}. We assume the attacker knows the reputation and allocation mechanisms being used and can execute an optimal attack.

\begin{definition}[Sybil Attack]
\label{def:sybilattack}
Given the feedback graph $G=(V,E,w)$, an attacker $s_0$ performs a Sybil attack $\sigma_S$ by introducing the following elements to the feedback graph: 
\begin{itemize}
	\item A set of \textbf{Sybil nodes} $S=\left\lbrace{} s_{1},\ldots{},s_{m}\right\rbrace \cup \lbrace{}s_{0}\rbrace$, each of which is indistinguishable from an honest node by other nodes.
	\item A set of \textbf{Sybil edges} $E_S\subset{}S\times{}S$ with arbitrary edge weights $w_S:S\times S\rightarrow{}\mathbb{R}_{\geq{}0}$.
	\item A set of \textbf{attack edges} $E_{a}\subset{}V\times{}S$ with weights $w_{a}:V\times S\rightarrow\mathbb{R}_{\geq{}0}$.
\end{itemize}  

After an attack has been carried out, we obtain a modified feedback graph, denoted by $G' := (V \cup S, E \cup E_S \cup E_a, w \cup w_S \cup w_a)$. We denote by $G'' := (V \cup S, E \cup E_a, w \cup w_a)$ the modified feedback graph with Sybil edges removed. 
\end{definition}

\begin{figure*}[t]
\centering
\begin{subfigure}{0.46\textwidth}
\centering
\includegraphics[width=1.0\textwidth]{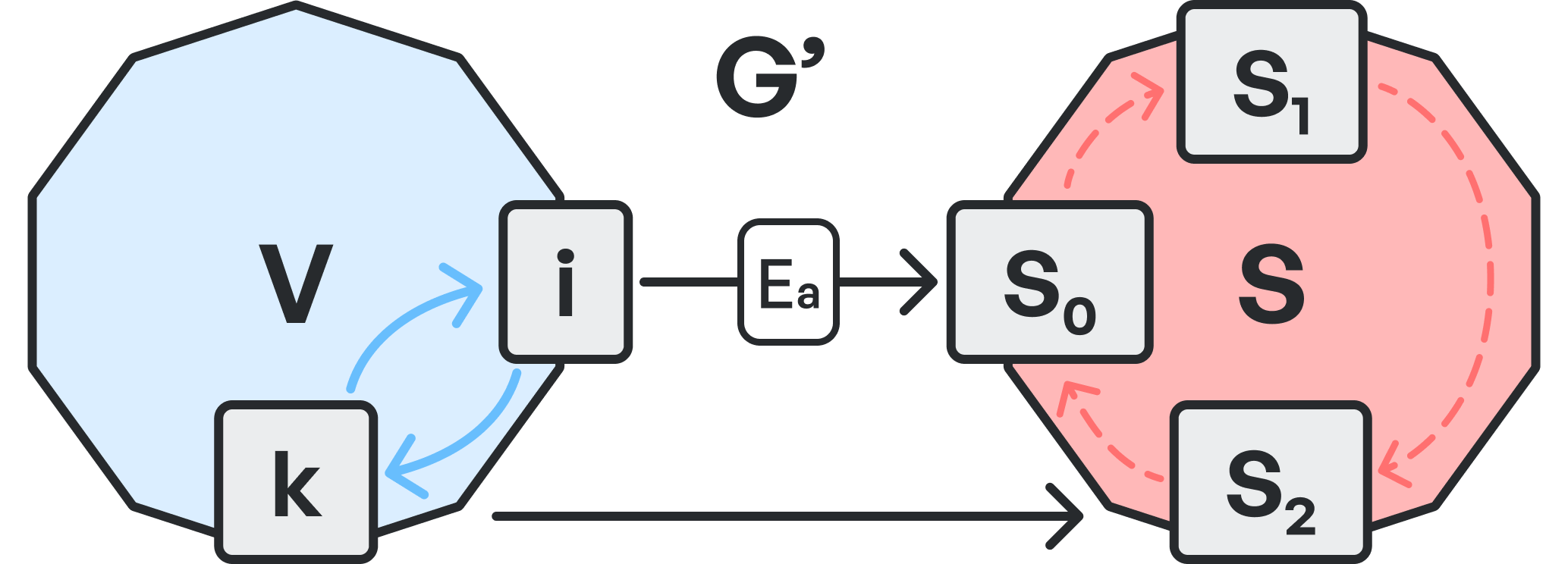}
\caption{The modified graph $G'$ after the attack.}\label{fig:modified_sybil_graph}
\end{subfigure} \quad
\begin{subfigure}{0.44\textwidth}
\centering
\includegraphics[width=1.0\textwidth]{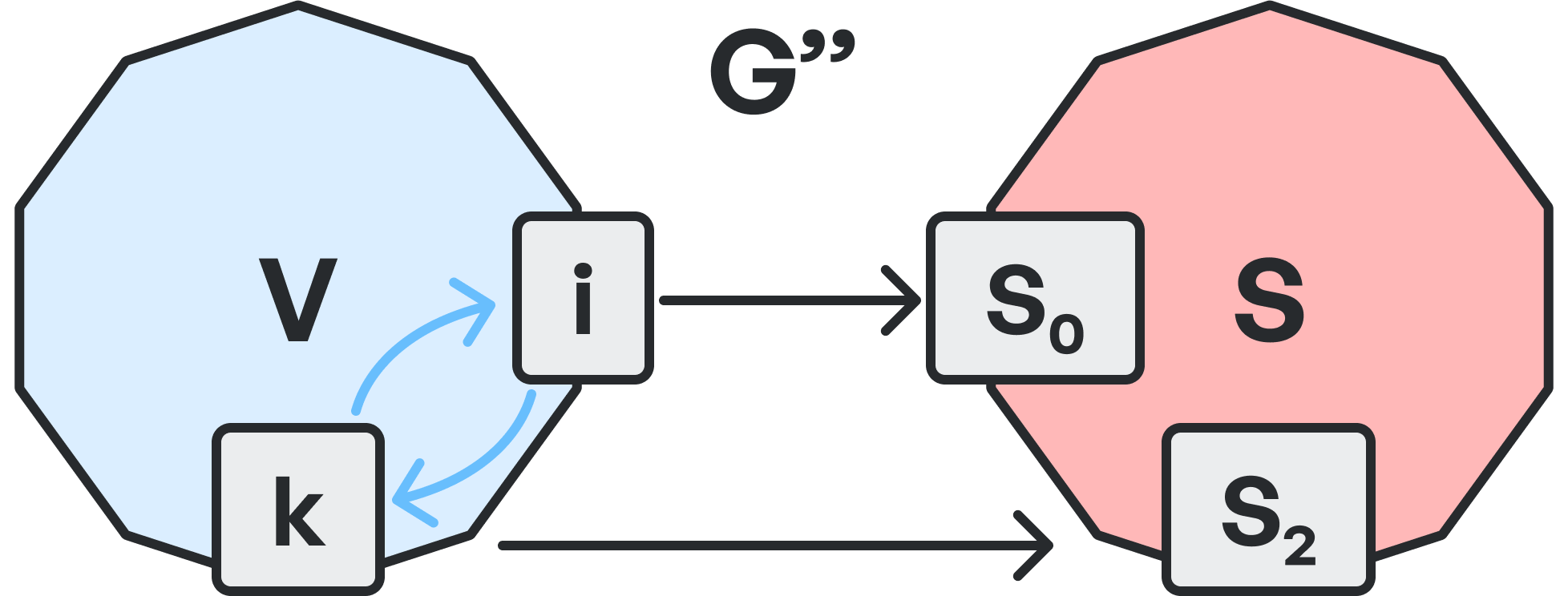}
\caption{The modified graph $G''$ with Sybil edges removed.}\label{fig:collapsed_sybil_graph}
\end{subfigure}\quad
\caption{An example of a Sybil attack on graph $G = (V=\{k,i,...\}, E=\{(k,i), (i,k), ...\}, w)$ with Sybil nodes $S = \{s_0, s_1, s_2\}$, two attack edges $E_a = \{(i,s_0),(k,s_2)\}$ and Sybil edges $E_S = \{(s_0,s_1), (s1,s2), (s2,s_0)\}$.}
\label{fig:sybil_attacks_example}
\end{figure*}

Figure~\ref{fig:sybil_attacks_example} illustrates an example of a Sybil attack. The attack has three Sybil nodes $\{s_0, s_1, s_2\}$. The attacker establishes Sybil edges between~$s_0$,~$s_1$, and~$s_2$ with arbitrarily high weights, as these edges can be freely generated by the attacker without any constraints. In contrast, attack edges typically require the attacker to make some real contributions to receive feedback from honest reputable nodes. To infiltrate the network, the attacker using identities of $s_0$ and $s_2$ performs some contribution to receive feedback from some highly reputable nodes~$i$ and~$k$ in~$V$, resulting in the creation of attack edges~$(i, s_0)$ and~$(k, s_2)$. As a result of this attack, the Sybil node~$s_1$ receives positive reputation. The created cycle $(s_0, s_1), (s_1, s_2), (s_2, s_0)$ with high weights further increases the reputation of each Sybil node, thereby inflating their reputation scores and allowing them to disproportionately benefit during the allocation phase.

\subsection{Sybil Attack Strategies}~\label{sec:exist_sybil_tolerance}

Naive approaches to reputation systems that rely on basic global centrality measures, such as degree centrality, are highly susceptible to manipulation. These approaches assign importance based on the number of direct connections a node has, making it easy for an attacker to inflate their reputation simply by creating Sybil nodes and edges that connect them to the attacker node. This artificially increases the attacker node’s degree centrality, thereby unfairly boosting its reputation.

More complex global centrality measures, such as (weighted) PageRank, are also vulnerable to Sybil attacks~\cite{cheng2005sybilproof}. An attacker can create a dense Sybil region, effectively increasing the PageRank of  Sybil nodes, even if the Sybil nodes themselves have low individual ranks. Moreover, an attacker can perform successful Sybil attacks without the need to create any attack edges.

We consider more sophisticated attack that require the creation of attack edges. Typically, an attacker infiltrates the system by interacting with honest nodes, gradually collecting reputation through before executing a Sybil attack. Once sufficient reputation is gained, the Sybil attack is launched, and some inflated reputation is gained. To further amplify the effect, we also consider a \textit{repeated Sybil attack}, where the attacker introduces a fresh batch of Sybil nodes in every epoch. This strategy compounds the attack’s impact over time, as each new set of Sybil nodes reinforces the influence of the malicious node through an expanding web of attack edges.

Personalized reputation mechanisms~\cite{liu2016personalized} inherently limit the scope of Sybil attacks by tying reputation to connectivity with seed nodes. A node can only gain a positive reputation if it is connected to a seed node through some path or directly. In systems with multiple seed nodes the challenge for attackers increases significantly. To inflate the reputation of a malicious node in such systems, the attacker must establish paths to each relevant seed node. If no such paths are established, the malicious node’s reputation score remains zero, effectively neutralizing the potential for Sybil attacks.

\begin{figure*}
\centering
\begin{subfigure}{0.31\textwidth}
\centering
\includegraphics[width=0.63\textwidth]{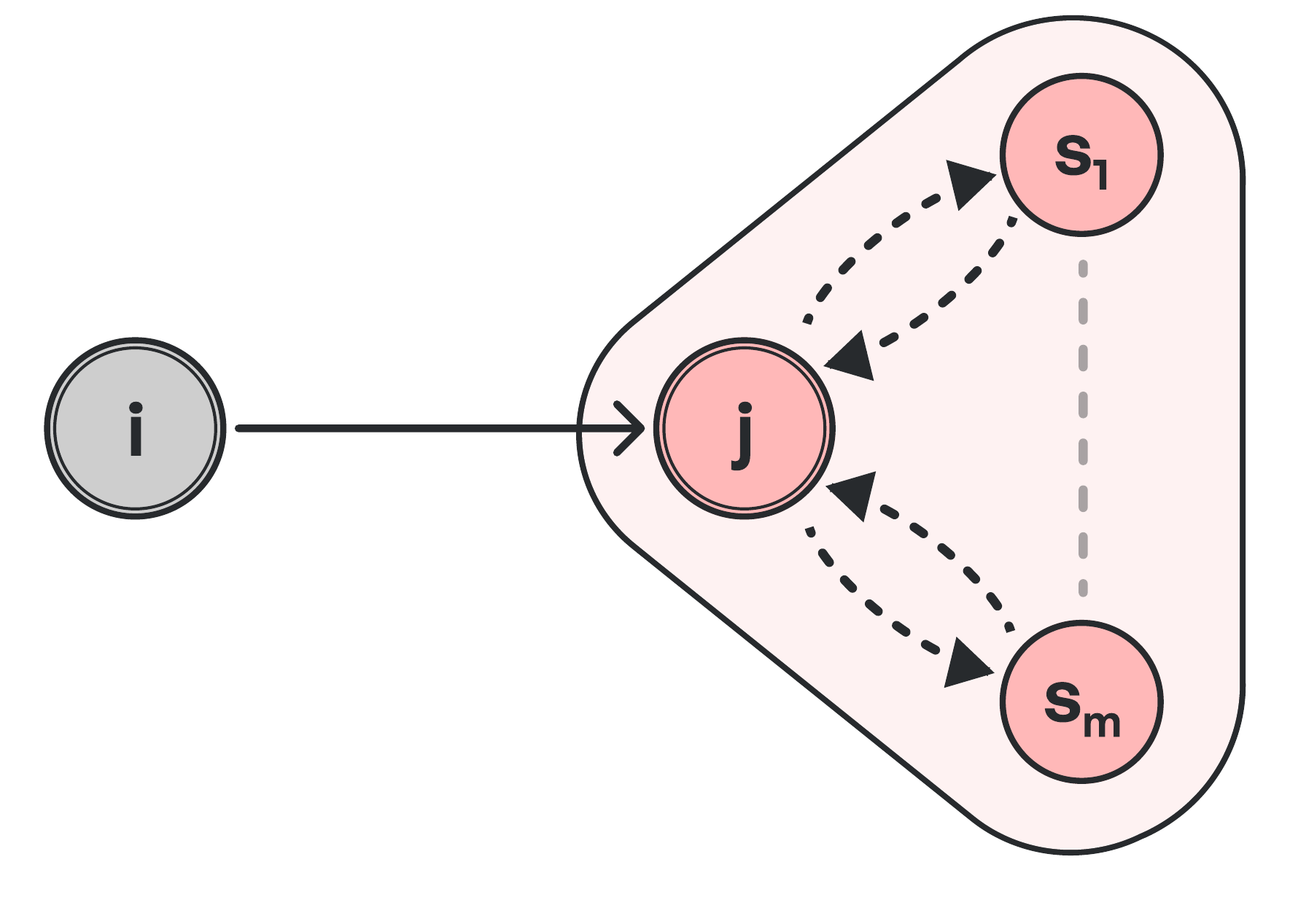}
\caption{Cycle attack}\label{fig:cycle_attack}
\end{subfigure} \quad
\begin{subfigure}{0.33\textwidth}
\centering
\includegraphics[width=1.0\textwidth]{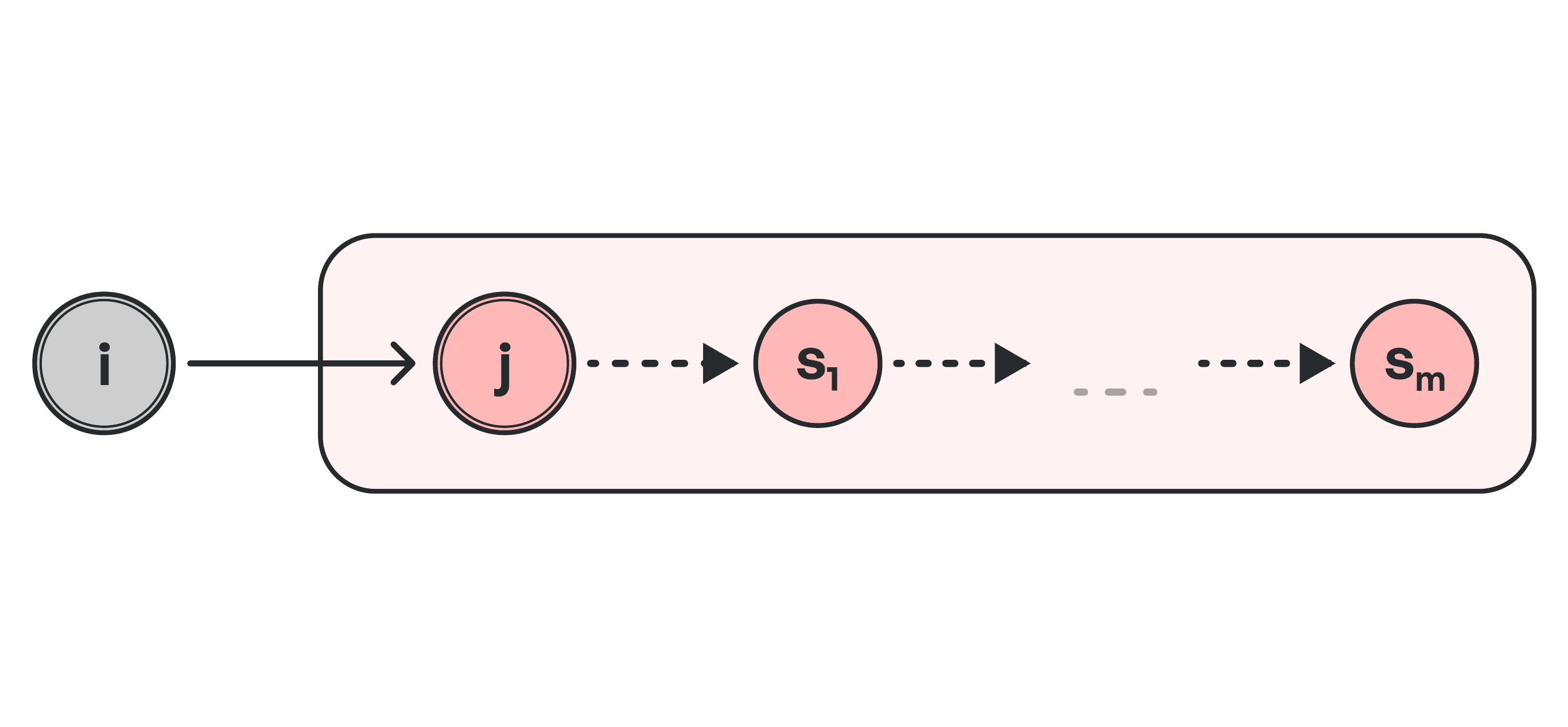}
\caption{Serial attack}\label{fig:path_attack}
\end{subfigure}\quad
\begin{subfigure}{0.31\textwidth}
\centering
\includegraphics[width=0.63\textwidth]{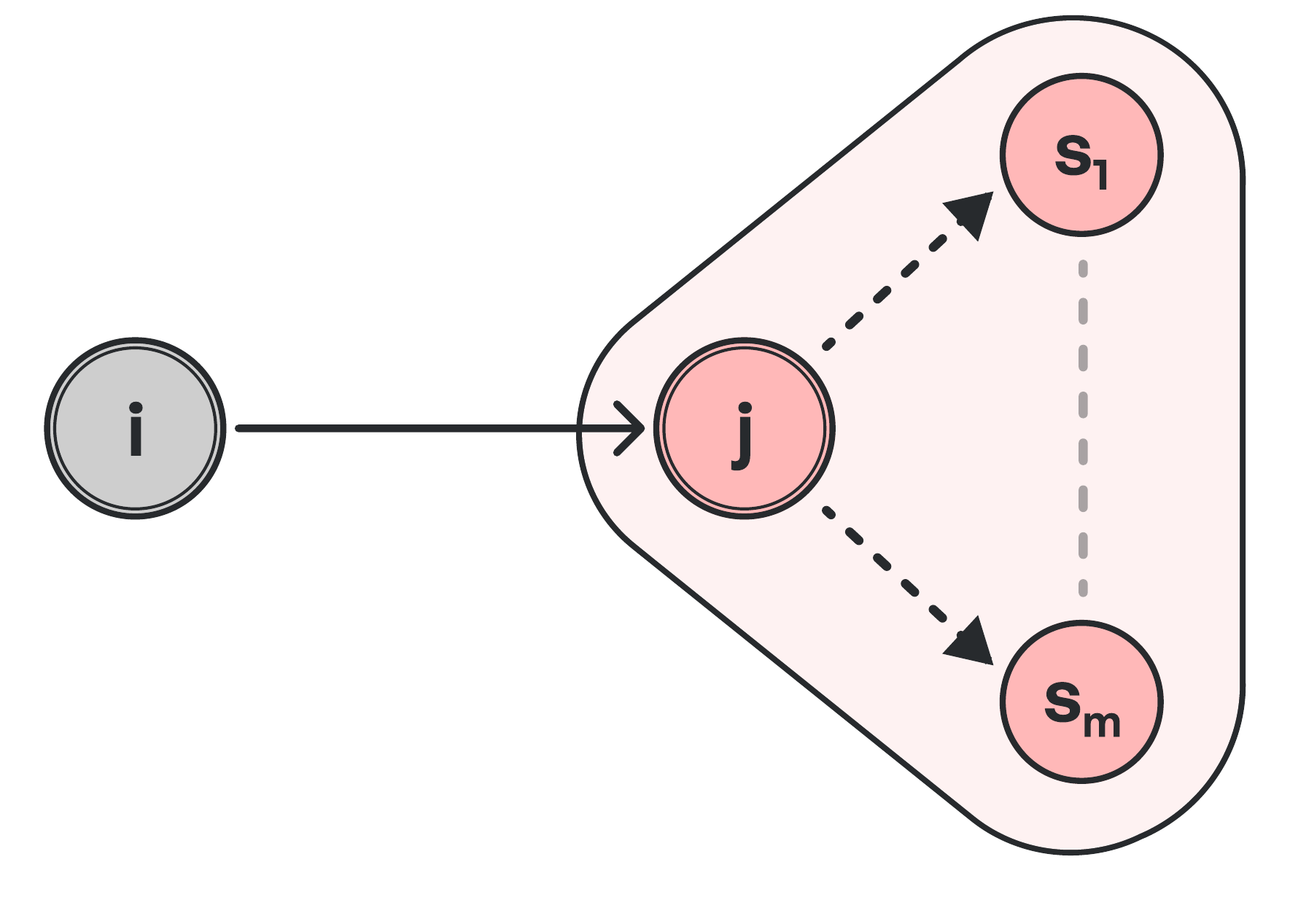}
\caption{Parallel attack}\label{fig:parallel_attack}
\end{subfigure}\quad
\caption{Sybil attack strategies. A beneficial Sybil attack is a combination of these three strategies.}
\label{fig:attacks}
\end{figure*}

We now present three Sybil attack strategies, which, when combined, cover all Sybil attack strategies as shown in the previous works~\cite{sybil_proofness_2021}. These strategies are illustrated in Figure~\ref{fig:attacks}.

\begin{definition}[Sybil Attack Strategies]
\label{def:attackstrategies}
An attacker node $s_0$ initially creates an attack edge $(i, s_0)$ with some node $i \in V$ by making a contribution and receiving a feedback from node $i$. Subsequently, it creates $m$ Sybil identities and Sybil edges, resulting in the modified feedback graph $G'(m)$. The Sybil edges are created in one of the following ways:
\begin{itemize}
\item \textbf{Cycle attack}. The attacker node $s_0$ creates both incoming and outgoing edges with each other Sybil node, forming edges $(s_0, s_k)$ and $(s_k, s_0)$ for $k \in \{1, \ldots, m\}$. 
\item \textbf{Serial attack}. The attacker node $s_0$ creates a linear sequence of edges $(s_0, s_1)$, and $(s_k, s_{k+1})$ for  $k \in \{1, \ldots, m-1\}$.
\item \textbf{Parallel attack}. The attacker node $s_0$ creates directed outgoing edges to each other Sybil node, forming edges $(s_0, s_k)$ for  $k \in \{1, \ldots, m\}$.
\end{itemize}
\end{definition}

\noindent 	We refer to the modified feedback graph  $G{\prime}(m)$  as the graph obtained after a Sybil attack with exactly  m  Sybil nodes and the corresponding Sybil edges added according to one of the attack strategies defined above.

\subsection{Sybil Tolerance}\label{sec:sybil-tolerance}

A \emph{beneficial} Sybil attack occurs when an attacker successfully inflates the reputation of Sybil nodes, allowing them to receive a disproportionate share of rewards. The attacker is allowed to create an arbitrary number of Sybil identities and establish fake edges between them to boost their reputation within the Sybil network. However, any edge between honest nodes in $V$ and Sybils in $S$ must represent a real transaction. This means that the attacker must make genuine contributions to receive feedback from honest nodes, thereby creating legitimate directed edges connecting the honest network to the Sybil identities. To maximize their gains, the attacker might target highly reputable nodes to create these legitimate attack edges, as feedback from such nodes would significantly boost the reputation of the connected Sybil nodes.

We distinguish between two types of reputations in this context: \textit{deserved} and \textit{inflated}. The \textit{deserved} reputation represents the value earned through legitimate feedback provided by honest nodes. In contrast, the \textit{inflated} reputation includes both the deserved reputation and any additional reputation artificially increased through fake edges between Sybil identities. The goal of the attacker is to inflate the reputation as much as possible, maximizing the cumulative reputation of the Sybil nodes through these fake edges.

\begin{superdefinition}
  \begin{subdefinition}{Inflated Reputation}
  Given an arbitrary seed node \( i \in V \), the inflated reputation \( \omega^{+}(\sigma_S) \) for a Sybil attack \( \sigma_S \) is the cumulative reputation score gained by the Sybil nodes in \( S \) calculated by \( i \) over the modified feedback graph \( G' \): 
  \begin{equation*}
  \omega^{+}(\sigma_S) = \sum\limits_{s \in S} R_i(G'_i, s)
  \end{equation*}
  \end{subdefinition}

  \begin{subdefinition}{Deserved Reputation}
  Given an arbitrary seed node \( i \in V \), the deserved reputation \( \omega(\sigma_S) \) for a Sybil attack \( \sigma_S \) is the cumulative reputation score gained by the Sybil nodes in \( S \) calculated by \( i \) over the modified feedback graph \( G'' \) with the Sybil edges removed:
  \begin{equation*}
  \omega(\sigma_S) = \sum\limits_{s \in S} R_i(G''_i, s)
  \end{equation*}
  \end{subdefinition}

  \begin{subdefinition}{Attacker's Gain}
  The attacker’s gain of a Sybil attack given $\omega(\sigma_S) > 0$ is defined as the ratio of the inflated and deserved reputation:
  \begin{equation*}
  \text{Gain}(\sigma_S) = \frac{\omega^{+}(\sigma_S)}{\omega(\sigma_S)}
  \end{equation*}
  \end{subdefinition}
\end{superdefinition}

Sybil tolerance is modeled as an upper bound on the attacker’s gain for a Sybil attack, denoted as $\sigma_S$, on a feedback graph $G$. We formally define a Sybil-Tolerant reputation mechanism as follows:

\begin{definition}[Sybil Tolerance]
\label{def:tolerance}
A reputation mechanism is \emph{Sybil-Tolerant} if, for every Sybil attack, the attacker’s gain is bounded by some constant $c \geq 1$ for arbitrarily large sets $S$ of Sybil nodes with any arrangement of Sybil edges, that is, if
\begin{equation*}
    \lim\limits_{|S|\to \infty}\frac{\omega^{+}(\sigma_S)}{\omega(\sigma_S)} \leq c
\end{equation*}
\end{definition}

In theory, a Sybil-resistant mechanism is characterized by having $c = 1$, meaning the attacker cannot gain any inflated reputation through Sybil identities and can only earn what is deserved from feedback provided by honest nodes. However, in practice, achieving full Sybil resistance $( c = 1 )$ in open, decentralized systems is extremely challenging due to the ease of creating pseudonymous identities and the potential for Sybil nodes to receive legitimate feedback from honest participants. Therefore, while mechanisms like restricting nodes to self-assessments can theoretically achieve $c = 1$, they severely limit the utility of the reputation system by ignoring valuable indirect feedback and collaboration.A notable example of this type of restrictive mechanism is the use of direct reciprocity strategies, such as Tit-for-Tat~\cite{nowak_five_2006}. In the Tit-for-Tat strategy, nodes reciprocate interactions, meaning they only provide positive feedback to nodes that have previously contributed to them in some way. This approach, widely known from its use in file-sharing protocols like BitTorrent~\cite{bittorent}, ensures that nodes cooperate only if they receive direct value from the interaction.

On the other hand, if the feedback from all peers is taken into account, Sybil tolerance is not achievable, as reported by Seuken and Parkes~\cite{seuken2014work}. The attacker's gain from a Sybil attack can become unbounded by creating a large Sybil region and amplifying the weights of Sybil edges. In practice, this means that the attacker can effectively receive a substantial reward for each allocation.

\subsection{Sybil-Tolerance of Existing Reputations}\label{sec:existing_reputation}

We evaluate three widely-used reputation mechanisms considered to be Sybil-resistant: personalized PageRank~\cite{bahmani2010pagerank}, personalized Hitting Time~\cite{liu2016personalized}, and BarterCast MaxFlow~\cite{meulpolder2009bartercast}. In this section, we present examples demonstrating that these mechanisms are not Sybil-resistant and, in some cases, fail to even achieve Sybil-tolerance.

\textbf{Personalized PageRank:} The personalized PageRank algorithm (PPR) is a variant of the standard PageRank algorithm, tailored to measure the importance of nodes relative to a specific seed node $i$. The algorithm works as follows: A random walk is initiated from the seed node $i$. At each step, with probability $\alpha$, the walk terminates and `teleports' back to the seed node $i$, and with probability $1-\alpha$, the walk moves to a randomly selected neighboring node. The PPR reputation of a node $j$ is the steady-state probability that the random walk initiated at $i$ will be at $j$ at any given step. In other words, it represents the probability that a random walker starting from the seed node $i$ will be found at node $j$ after any number of steps, assuming the process has reached equilibrium. In practice, this is estimated by running multiple random walks starting from the seed node $i$ and counting the number of times node $j$ is reached.

One significant issue with PageRank is that random walks can become trapped in the Sybil region, especially when a cycle attack is executed. In the global version of PageRank, this can occur even without the presence of attack edges. The most effective method for attacking PageRank is through a cycle attack, where the attacker creates edges with large weights among Sybil nodes. This manipulation increases the total number of encounters in a random walk, thereby enabling unbounded inflation of the reputation scores of the Sybil nodes and making PageRank not Sybil-tolerant.

PPR introduces a level of Sybil tolerance due to its damping factor  $\alpha$. This damping factor ensures that at each step, there is a probability $\alpha$ of returning to the seed node, which reduces the likelihood of random walks becoming trapped in the Sybil region. PPR was previously reported as Sybil-resistant, but this is true only in the absence of attack edges, where no random walks would reach any of the Sybil nodes. However, when attack edges are present, the attacker can still obtain a gain from a Sybil attack, especially when $\alpha$ is low.

\textbf{Personalized Hitting Time:}
The Personalized Hitting Time (PHT) algorithm is a variation of the random walk-based algorithms. Unlike PPR,  which focuses on steady-state probabilities and the long-term presence of nodes, PHT concentrates on the first encounter of nodes during the random walk. It measures the expected number of steps a random walker, starting from a seed node $i$, will take to reach target node $j$ for the first time.

Compared to PageRank, PHT is less affected by cycle attacks. However, PHT remains susceptible to serial attacks, where Sybil identities are connected in a linear sequence. In this scenario, each Sybil node in the sequence gains some positive reputation as the random walk progresses through the chain, ultimately leading to the inflation of the reputations of all Sybil nodes along the path. Similar to PPR, PHT is not inherently Sybil-resistant but can achieve some level of Sybil tolerance by using a high teleportation probability $\alpha$.

\textbf{MaxFlow:} The MaxFlow (MFW) algorithm is a classical approach in network flow theory that determines the maximum possible flow from a source node to a target node in a flow network. Given a directed graph $G = (V, E, w)$ where each edge $(i, j) \in E$ has a capacity $w \geq 0$, the objective is to find the maximum flow from a source node $i$ to a sink node $j$ such that the flow on each edge does not exceed its capacity and the incoming flow equals the outgoing flow for every node except $i$ and $j$. 

MFW was proposed for use in reputation systems because it effectively mimics the idea of credit networks~\cite{viswanath2012canal}, where the flow of value or contributions can be tracked through paths in a network. In such systems, the reputation of a node is determined by its ability to contribute value that can reach the target node, considering all possible routes. This approach ensures that the amount a node can gain is bounded by its aggregated contributions, making it a robust measure of a node’s overall influence and trustworthiness within the network.

Despite its theoretical benefits, MFW-based reputation mechanisms are vulnerable to Sybil attacks due to how they aggregate flow across multiple paths. In a parallel attack (shown in Figure~\ref{fig:parallel_attack}), an attacker creates multiple Sybil identities, each connected directly to the attacker’s main node $s_0$. Since MFW sums the flow through all paths, each additional Sybil identity increases the total flow capacity from honest nodes to the attacker, allowing the attacker to artificially inflate its cumulative reputation. This results in an unbounded gain because there is no constraint on the number of Sybil identities an attacker can create.

\subsection{Bounding Properties for Sybil-Tolerance}\label{sec:properties}

In this section we define three desirable properties of reputation mechanisms to achieve the Sybil-Tolerance.  

The first property we define aims to bound the effectiveness of parallel and cycle Sybil attacks. These attacks are effective because reputation mechanisms typically treat each new edge as an independent source of feedback, resulting in an inflated reputation for the attacker. To address this issue, we define a property called the \textit{Parallel Attack Bound}. This property ensures that, regardless of the number of Sybil nodes introduced, the cumulative reputation gain remains bounded, so the attacker’s inflated reputation does not exceed what would be achieved by a single Sybil node.

\begin{definition}[Parallel Attack Bound]~\label{def:parallel_attack_bound} A reputation mechanism is parallel attack bound if, for any seed node $i \in V$ and for the set of Sybil nodes $s_1,s_2,...,s_m$  generated through a parallel or cycle attack, the total reputation satisfies:
\begin{equation*} 
\sum_{l=1}^{m} R_i(G_i'(m), s_l) \leq R_i(G_i'(1), s_1).
\end{equation*}
\end{definition}

The second property we define is aimed at bounding the effectiveness of serial Sybil attacks. In such attacks, the attacker extends the path of Sybil nodes, with each new Sybil contributing additional feedback along the chain, thereby inflating the attacker’s reputation. To address this, we introduce the \textit{Serial Attack Bound}. This property ensures that the inflated reputation of a serial attack is bounded, even as the length of the Sybil paths increases, preventing the attacker from achieving unbounded gains by simply extending the chain of Sybil nodes.

\begin{definition}[Serial Attack Bound]~\label{def:serial_attack_bound} A reputation mechanism is serial attack bound if, for any seed node $i \in V$ and for the set of Sybil nodes $s_1, s_2, ..., s_m$ generated through a serial attack, the total reputation satisfies:
\begin{equation*} 
\lim\limits_{m \to \infty} \sum_{l=1}^{m} R_i(G_i'(m), s_l) < \infty. 
\end{equation*} \end{definition}

The \textit{Bounded Transitivity} property ensures that reputation values cannot be inflated through paths that include low-reputation nodes. Our proposed decay mechanisms, particularly transitivity decay, directly enforce this property by reducing the influence of nodes as their distance from the seed node increases. This means that reputation cannot be disproportionately amplified through long chains of intermediaries, which often include low-reputation or Sybil nodes. By integrating decay mechanisms, we ensure that the reputation system adheres to Bounded Transitivity, enhancing its resistance to manipulation. This principle has been explored in reputation algorithms that incorporate MaxFlow concepts, such as BarterCast~\cite{meulpolder2009bartercast}.

\begin{definition}[Bounded Transitivity]\label{def:bounded_transitivity}
A reputation mechanism satisfies \emph{Bounded Transitivity} if, for any seed node \( i \in V \) and any node \( j \in V_i \) such that there are \( N \) node-disjoint, simple paths \( P_d \) (where \( d = 1, 2, \dots, N \)) from \( i \) to \( j \), it holds that:
\[
R_i(G_i, j) \leq \sum_{d=1}^{N} \min \{ R_i(G_i, k) \mid k \in P_d \}.
\]
\end{definition}

\section{MeritRank: Sybil Tolerant Reputations}\label{sec:meritrank}

In this section, we introduce \meritrank{}, a set of techniques designed to enhance the Sybil tolerance of existing reputation mechanisms. 

\subsection{Bounding Properties in Practice}

To achieve the bounding properties discussed in the Section~\ref{sec:properties} in practice, we propose four generic techniques for existing reputation mechanisms. Specifically, we introduce relative feedback, decay on transitivity, and decay on connectivity as heuristics to achieve bounds on Sybil attacks. Additionally, we consider epoch decay due to its popularity in existing reputation mechanisms.

\textbf{Relative Feedback}. In a Sybil attack, the attacker can control the edge weights of Sybil nodes, potentially assigning arbitrarily large values to inflate the influence of these connections. Without proper normalization, these artificially high weights can distort the reputation mechanism, as the system might rely excessively on the absolute values of the incoming edges. To counteract this, Relative Feedback introduces a normalization step to each edge weight, ensuring that no single node can amplify its reputation through artificially inflated connections.

We introduce the Normalized Graph $G^{N}$ of a weighted graph $G = (V, E, w)$ in the following way: for every node \( k \in V \), let \( \mathcal{N}(k) \) be the set of its neighboring nodes in \( G \). The corresponding \emph{Normalized Graph} \( G^{N} = (V, E, w^{N}) \) is defined such that:

\begin{equation*}
w^{N}(k, j) = \frac{w(k,j)}{\sum\limits_{l\in \mathcal{N}(k)} w(k, l)}, \quad \forall (k, j) \in E.
\end{equation*}

The normalization process is applied to every subjective graph, adjusting all edge weights accordingly. For flow-based algorithms, this normalization is applied prior to calculating the reputation scores. For random-walk based mechanisms, the normalization is incorporated dynamically during the calculation of the transition probabilities, with each step reflecting the probability of moving from node $k$  to node  $j$  based on  $w^{N}(k, j)$. 

This mechanism is the key to achieve Bounded Transitivity property and contributes to the Parallel Attack Bound. It ensures that the reputation is always considered in the context of its neighbors. Attackers cannot inflate reputations by assigning large weights to edges connected to Sybil nodes, as the normalization limits their impact.

\textbf{Transitivity Decay}. The purpose of this decay is to decrease the reputation of nodes as their distance from the seed node increases. This discourages the creation of long chains, such as those used in serial attacks, by ensuring that nodes further from the seed receive progressively lower scores.

In random-walk-based mechanisms, this concept is naturally implemented using a teleportation probability $\alpha$. With a probability of $\alpha$, a random walk returns to the seed node, limiting the influence of nodes further down the path. Typically, 
$\alpha$ is set between 0.1 and 0.2~\cite{pagerank1998}. While we reuse this idea, we reinterpret $\alpha$ not merely as a teleportation probability but as a decay factor that systematically reduces a node’s reputation contribution based on its distance from the seed node. Instead of merely terminating random walks, $\alpha$ serves to attenuate the influence of nodes further away, ensuring that the reputation impact diminishes with each additional hop in the network.

In a flow-based reputation system, transitivity decay is implemented by introducing a reduction factor applied to the flow. Specifically, the flow $w$ from node \(i\) to node \(j\) is modified to \(w'(e) = (1 - \alpha)^d \times w(e) \), where \(d\) is the distance from the seed node to node \(j\). 

Transitivity Decay directly enforces the Serial Attack Bound by diminishing the influence of nodes further from the seed node, thus limiting the cumulative reputation that can be gained through serial attacks.

\textbf{Connectivity Decay}. Sybil nodes often form dense clusters that connect to the rest of the network through a small number of intermediary nodes known as \emph{cut-vertices}. A \emph{cut-vertex} is a node whose removal increases the number of connected components in a graph, potentially isolating certain nodes. Identifying and accounting for cut-vertices is essential in mitigating the influence of nodes that depend on such vulnerable connections.

The \emph{connectivity decay} mechanism adjusts the reputation scores of nodes that depend on cut-vertices, thereby reducing the influence of potential Sybil clusters. The adjusted reputation score with connectivity decay for a node \( j \in V_i\) with respect to a seed node \( i \), is defined as:

\begin{equation*}
R_{i,\beta}(G_i, j) =
\begin{cases}
(1 - \beta) \cdot R_i(G_i, j), & \text{if } I(i, j) = 1, \\
R_i(G_i, j), & \text{otherwise},
\end{cases}    
\end{equation*} 

\noindent where \( 0 \leq \beta \leq 1 \) is the decay factor, and \( I(i, j) \) is an indicator function that determines whether the connection between nodes \( i \) and \( j \) is considered vulnerable due to low \emph{local node connectivity}.

The concept of \emph{local node connectivity} between two nodes \( i \) and \( j \), denoted \( \kappa(i, j) \), is defined as the minimum number of nodes (excluding \( i \) and \( j \)) whose removal would disconnect \( i \) from \( j \). Equivalently, \( \kappa(i, j) \) is the maximum number of \emph{node-independent} paths between \( i \) and \( j \), where node-independent paths are paths that do not share any nodes other than the endpoints. We introduce a \textit{connectivity threshold} \( t \) (with \( t \geq 1 \)) to determine the acceptable level of connectivity between nodes. If the local node connectivity \( \kappa(i, j) \) is less than or equal to \( t \), the connection between \( i \) and \( j \) is considered vulnerable and subject to decay. The indicator function \( I(i, j) \) is then defined as:

\begin{equation*}
I(i, j) =
\begin{cases}
1, & \text{if } \kappa(i, j) \leq t, \\
0, & \text{otherwise}.
\end{cases}    
\end{equation*}

In practice, calculating $\kappa(i, j)$ directly can be computationally intensive for large networks. To address this, we perform a large number of random walks initiated from a seed node, with each node calculating locally based on its own subjective feedback graph. The indicator function $I(i, j)$ is estimated by analyzing the frequency of random walks from node $i$ to node $j$ that pass through intermediary nodes. This estimation is practical because random walks are computationally efficient and can be performed locally by every node using its subjective view of the network.

Let $T_{ij}$ denote the total number of random walks from node $i$ to node $j$, and let $T_{ij}(k)$ represent the number of these walks that pass through an intermediary node $k \in V \setminus \{i, j\}$. A node $k$ is considered critical if the proportion of walks passing through $k$ exceeds a threshold of $1 / t$. Thus, the indicator function $I(i, j)$ is equivalently defined as:

\begin{equation*}
I(i, j) =
\begin{cases}
1, & \text{if } \displaystyle \max_{k \in V \setminus \{i, j\}} \left( \frac{T_{ij}(k)}{T_{ij}} \right) \geq \dfrac{1}{t}, \\
0, & \text{otherwise}.
\end{cases}    
\end{equation*}

This approximation relies on the idea that when \( \kappa(i, j) \) is low, there are fewer node-independent paths between \( i \) and \( j \), causing a higher proportion of random walks to pass through certain critical nodes. The threshold \( t \) determines the sensitivity of the connectivity decay mechanism. The higher $t$, the more node-independent paths are required to avoid the decay.

Connectivity decay penalizes nodes that are connected to the seed node through a small number of independent paths. In a parallel attack, an attacker creates multiple Sybil nodes directly connected to itself, which is then connected to the seed node through limited attack edges. By applying connectivity decay, the reputation influence of each Sybil node is reduced because they all rely on the same limited connections to the seed node, enforcing the Parallel Attack Bound. Similarly, connectivity decay impacts serial attacks by reducing the influence of Sybil nodes that are connected to the seed node through a single, elongated path, thereby achieving the Serial Attack Bound.

\textbf{Epoch Decay}. Attackers might attempt to exploit the system by maintaining high reputations for their Sybil nodes based on the older feedback. To counter this, we introduce the concept of \emph{epoch decay}, which reduces the influence of feedback as it ages. The epoch decay aims to require continuous positive contributions to maintain a high reputation. However, epoch decay might have unintended consequences. For instance, it could uniformly reduce the reputations of all participants and, as a result, potentially favor attackers who can rapidly generate new Sybil identities.

Given an epoch decay coefficient \( 0 \leq \gamma \leq 1 \), we detail two methods for implementing this decay:
\begin{enumerate}
    \item \textbf{Decay on Reputation Values}: 
    This method dynamically adjusts reputation values over time to give more weight to recent contributions. The reputation update formula for a node \( j \) from the perspective of a seed node \( i \) at epoch \( \tau + 1 \) is:
    \begin{equation*}
    R_i(G_i^{(\tau+1)}, j) = (1 - \gamma) \cdot R_i(G_i^{(\tau)}, j) + R_i(G_i^{\Delta(\tau+1, \tau)}, j)
    \end{equation*}

    \noindent Where, $G_i^{\tau} = (V_i^{\tau}, E_i^{\tau}, w_i^{\tau})$ is the subjective feedback graph of node $i$ at epoch $\tau$, and 
    \( G_i^{(\Delta(\tau+1, \tau))} = (V_i^{(\tau+1)}, \Delta E_i^{(\tau+1)}, \Delta w_i^{(\tau+1)}) \) represents the changes to the graph $G_i^{\tau}$ in epoch $\tau + 1$, which account for newly added edges and the weight updates for existing edges. 
    
    \item \textbf{Decay on Graph Weights}: This method applies decay directly to the weights of the graph’s edges, progressively decreasing the impact of older edges. The weight update formula for an edge $(i, j)$ at epoch $\tau+1$ is:
    \begin{equation*}
    w_{\gamma}^{(\tau+1)}(i, j) = \max\left(0, w^{(\tau+1)}(i, j) - (1 - \gamma) \cdot w^{(\tau)}(i, j)\right)
    \end{equation*}
\end{enumerate}

\section{Experiments} \label{sec:experiments}

In this section, we present the results of a quantitative study on the MeritRank algorithm. Our evaluation focuses on the impact of Sybil attacks on reputation scores and examines how the proposed decay mechanisms influence these scores. Additionally, we assess the effect of decay mechanisms on the informativeness of the reputation system, i.e., the system's ability to still accurately rank honest nodes.

\subsection{Experimental Set-Up}

For our study, we consider the MakerDAO forum, one of the largest decentralized autonomous organizations (DAOs) to date~\cite{makerdao2017whitepaper}. MakerDAO is a decentralized platform built on the Ethereum blockchain that facilitates the issuance and management of the \textit{DAI} stablecoin. The \textit{DAI} token is an algorithmically stabilized cryptocurrency pegged to the US dollar. MakerDAO is governed by holders of the \textit{MKR} utility token, which serves as the platform's governance token. \textit{MKR} token holders participate in decision-making processes by creating and voting on proposals within the MakerDAO forum~\cite{makerdaoforum}. These proposals influence MakerDAO’s operations, including adjustments to the parameters that maintain the stability of \textit{DAI}.

We prepare a dataset for our experiments by parsing the forum activity using the provided API\footnote{\url{https://forum.sky.money}}. Our dataset encompasses all user interactions within the MakerDAO forum, including replies, likes, posts, and votes on proposals, spanning from June 24, 2019, to May 26, 2022 (153 weeks). Each action within the forum is quantified in terms of \textit{work units}, following the framework defined by the SourceCred MakerDAO project\footnote{\url{http://makerdao.sourcecred.io}}. For example, a post is evaluated based on the number of likes it receives, with each like contributing four work units to the post's total value. Using these work units, we construct a work graph where nodes represent users and entities (such as posts), and edges represent actions connecting them. For instance, if a user creates a post and another user likes it, this interaction is represented in the graph as a sequence of edges: from the user to the post (creation) and from the post to the user who liked it (feedback).

We then derive a feedback graph by compressing the work graph. In this process, intermediary entities like posts and their associated edges are first aggregated into direct edges between users, which are then combined into a single edge. The result is a simplified graph where the edge weight $w(i,j)$ between two users $i$ and $j$ reflects the cumulative relative feedback given from user $i$ to user $j$. This edge weight is calculated by summing the work units of all  interactions—such as likes, replies, and other forms of feedback—that user $i$ has provided to the posts and comments made by user $j$.

We consider updates to the feedback graph over one-week epochs. Every epoch, the feedback graph is updated based on all forum interactions recorded that week. These updates appear as either new edges added between users or updated weights for existing edges. In total there are 153 epochs with peak activity occurring in week 149, when 1,528 new edges were added. By the end of epoch 153, the feedback graph comprised 2,057 nodes and 35,853 edges.

We create a simulation according to the model described in Section~\ref{sec:merit_model}. For consistency, we maintain a fixed seed node outside the graph. As the feedback graph evolves with new edges and updated weights, this seed node remains continuously connected to the top 10 currently most reputable nodes with equal weights. To evaluate~\meritrank{}, we implement two types of Sybil attacks: a \textit{single Sybil attack} and a \textit{repeated Sybil attack}.

Before initiating these attacks, we allow the system a 20-epoch grace period to establish initial reputations. At the end of this period, we designate one of the top 10 most reputable nodes as the attacker node. For the single Sybil attack, we simulate a scenario where an attacker infiltrates the system, executes a single attack with a varying number of Sybil nodes, and then withdraws, taking the profit.

In the repeated Sybil attack, the attacker consistently adds a new  Sybil node every epoch to the system after the grace period. Each new Sybil node added is a new node with no prior reputation. The  attacker node is selected once and fixed for the whole experiment run. This setup models a scenario where the attacker gradually increases its influence to poison the system and achieve larger cumulative gains. While less common in practice, this approach tests the system’s resilience under worst-case conditions.

\subsection{Transitivity Decay}

\begin{figure*}
	\centering
	\begin{subfigure}{.51\linewidth}
	    \centering
     \includegraphics[width=1\linewidth]{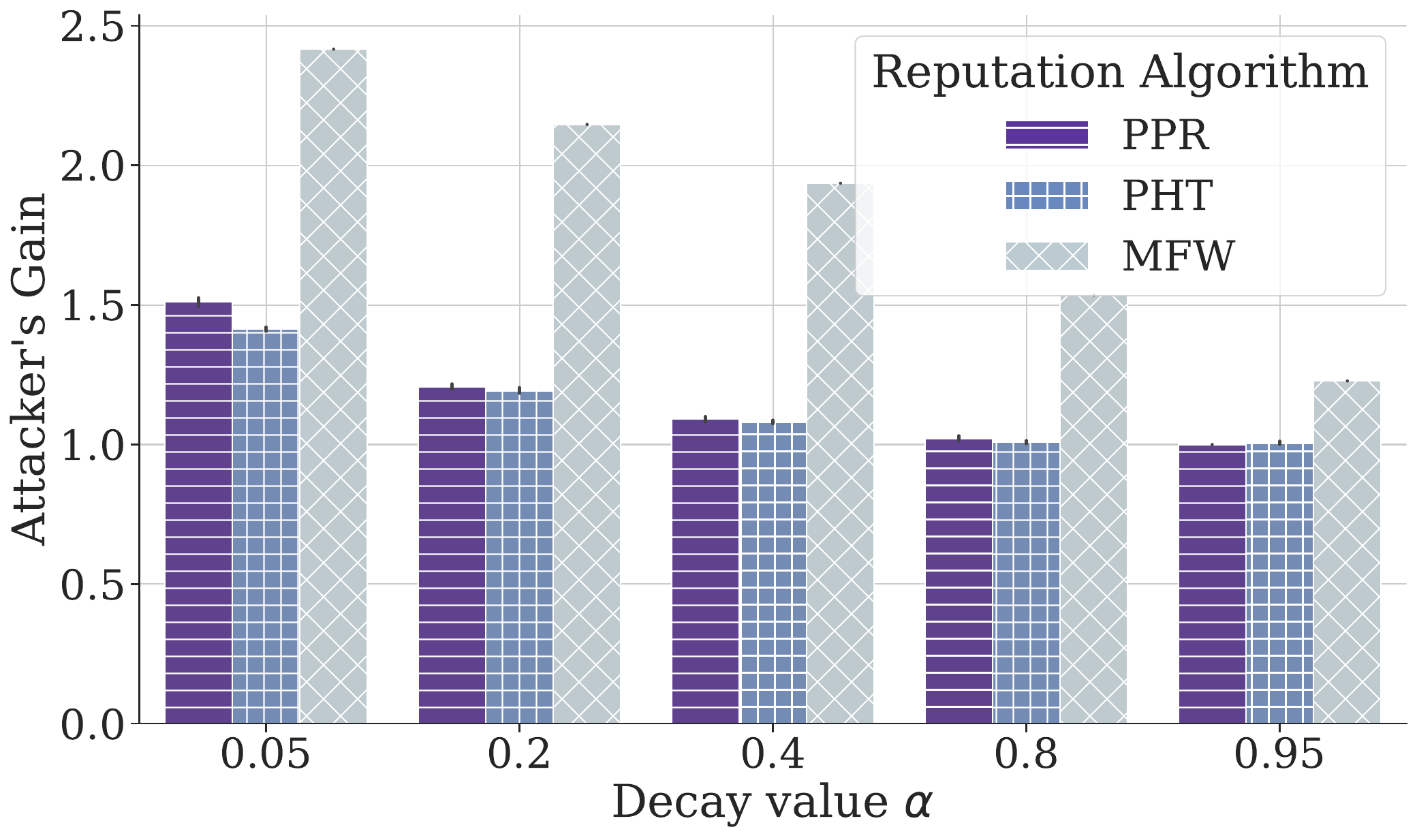}
	\caption{Attacker's gain with $50$ Sybil nodes.}
	\label{fig:alpha_bounds_single}
	\end{subfigure}
    \centering
    \begin{subfigure}{.47\linewidth}
	    \centering
    \includegraphics[width=1\linewidth]{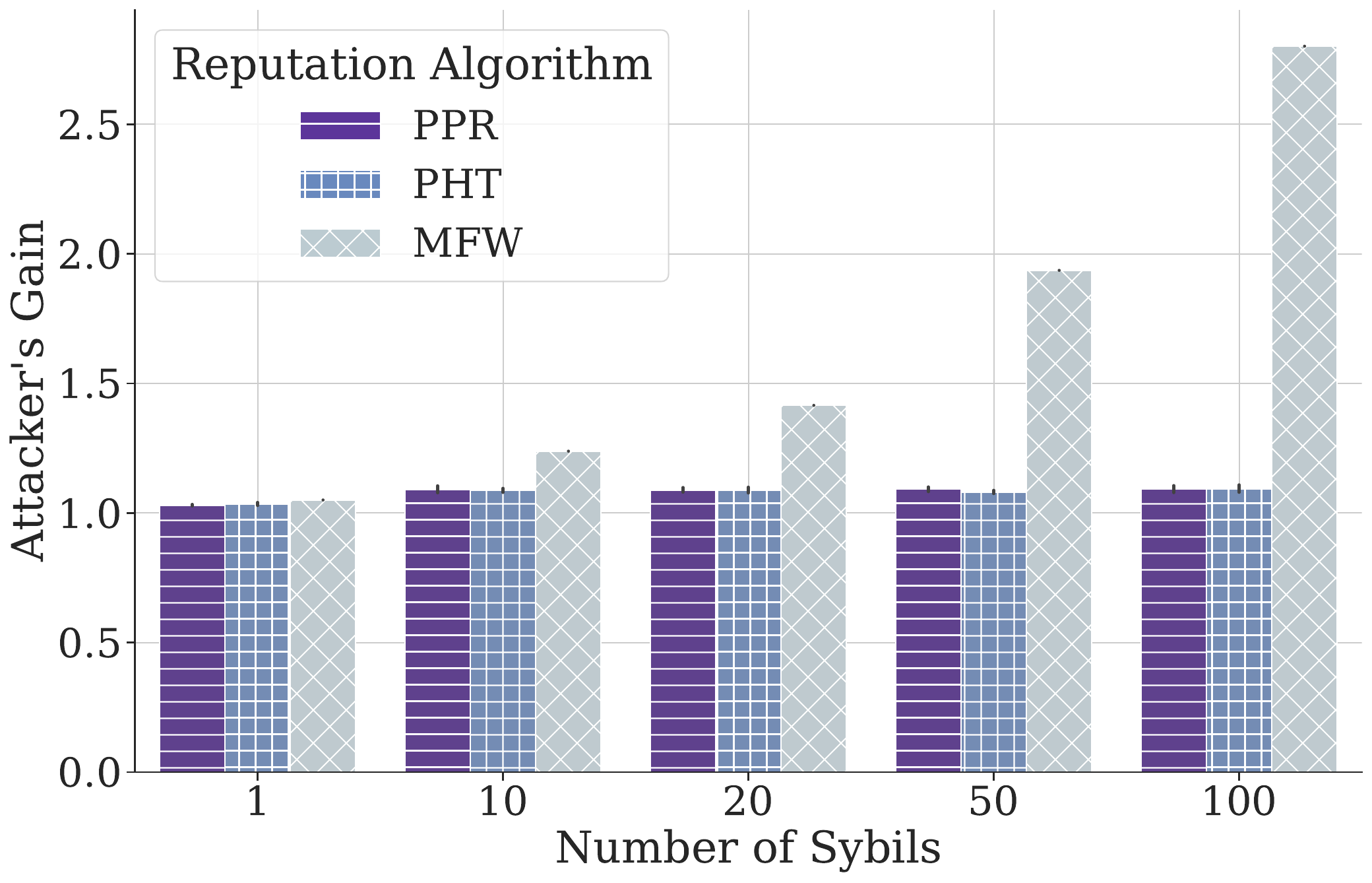}
    \caption{Attacker's gain given a transitivity decay value $\alpha$ of 0.4}
		\label{fig:alg_bounds_single}
	\end{subfigure}
	\caption{The attacker's gain from a single Sybil attack for the three reputation algorithms versus (a) the transitivity decay ($\alpha$) and (b) the number of Sybil nodes.}
	\label{fig:alpha_exp_single}
\end{figure*}

\begin{figure*}
	\centering
	\begin{subfigure}{.51\linewidth}
	    \centering
     \includegraphics[width=1\linewidth]{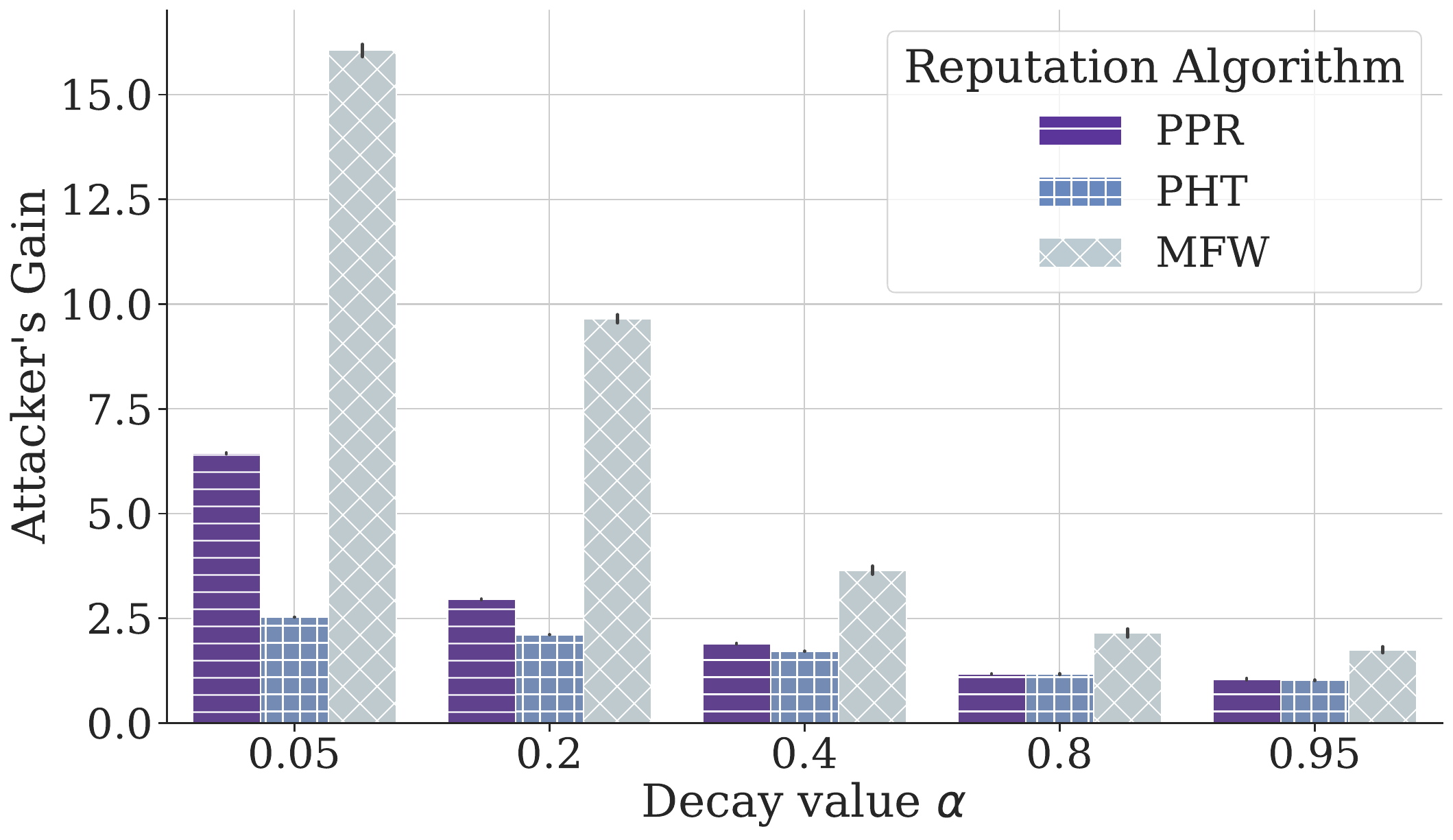}
	\caption{Attacker's gain with $50$ Sybil nodes added per epoch.}
	\label{fig:alpha_bounds}
	\end{subfigure}
    \centering
    \begin{subfigure}{.47\linewidth}
	    \centering
    \includegraphics[width=1\linewidth]{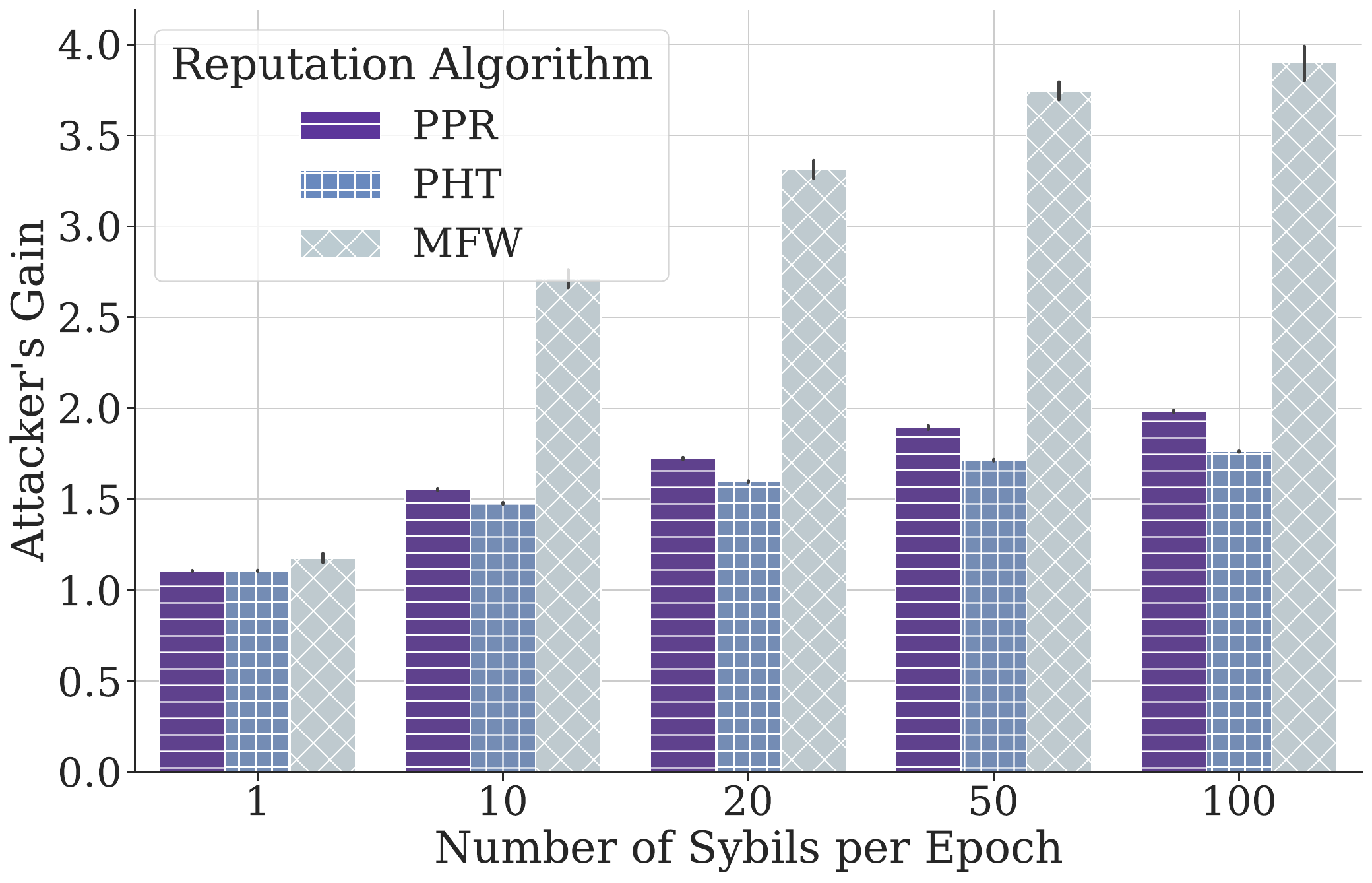}
    \caption{Attacker's gain given transitivity decay value $\alpha$ of 0.4}
		\label{fig:alg_bounds}
	\end{subfigure}
	\caption{The attacker's gain from a repeated Sybil attack for the three reputation algorithms versus (a) the transitivity decay ($\alpha$) and (b) the number of Sybil nodes per epoch.}
	\label{fig:alpha_exp}
\end{figure*}

We implement the Personalized PageRank (PPR), Personalized Hitting Time (PHT), and MaxFlow (MFW) algorithms, incorporating transitivity decay to assess the robustness of these reputation mechanisms against Sybil attacks. For each reputation mechanism, we introduce a varying number of Sybil nodes and corresponding Sybil edges, following the most effective attack strategy identified for each reputation in Section~\ref{sec:existing_reputation}. For all reported figures involving varying parameters, we repeat the simulation 10 times and present the average along with the standard error. Note that the standard error is not visible in the figures as its value is negligibly small.

We report the attacker’s gain from a single Sybil attack executed right after the grace period in Figure~\ref{fig:alpha_exp_single}. In Figure~\ref{fig:alpha_bounds_single}, we illustrate the effect of transitivity decay as a function of the decay value $\alpha$. We vary the decay value $\alpha$ from $0.05$, representing minimal transitivity decay, to $0.95$, where the most distant edges are severely decayed. Setting $\alpha = 0$ is not possible, as it would prevent random walks from terminating. We set the number of Sybil nodes to 50. The results show that the attacker can gain up to 1.5 to 2.5 times when there is minimal transitivity decay. However, as the decay value increases, the attacker’s gain decreases.

Figure~\ref{fig:alg_bounds_single} further examines the effect of increasing the number of Sybil nodes on the attacker’s gain. Even as the number of Sybil nodes increases, the attacker’s gain does not grow significantly except for MFW. Notably, the MFW mechanism shows a higher attacker’s gain compared to PPR and PHT.

To further analyze the cumulative impact of Sybil attacks over time, we execute repeated Sybil attacks and report the attacker’s gain at the end of 153 epochs. Our results are presented in Figure~\ref{fig:alpha_exp}. The repeated Sybil attack clearly shows that the attacker can gain more compared to the single Sybil attack, which is especially evident for MFW.

Our findings indicate that transitivity decay can successfully decrease the attacker’s gain, especially with higher decay values. Among the reputation mechanisms tested, PHT results in the lowest attacker’s gain for both single and repeated Sybil attacks. MFW has poor Sybil tolerance because it aggregates trust additively across all available paths, allowing attackers to inflate their reputation despite the decay. Moreover, MFW is computationally more intensive than PPR and PHT, making it less desirable in practice.

PPR performs worse than PHT because it is more vulnerable to cycle attacks. In PPR, reputation is distributed based on the stationary distribution of random walks with restarts, which can be influenced by the presence of cycles or loops in the network. Attackers can exploit this by creating cyclic structures among Sybil nodes to trap the random walks and thereby absorb more reputation. This results in higher reputation scores for Sybil nodes in PPR compared to PHT.

The intuition behind transitivity decay is that it is spatial—the further a node is from the seed node in terms of network distance, the more its reputation is decayed. Consequently, nodes that are farther away receive exponentially less reputation than those closer to the seed node. This mechanism is particularly effective against serial attacks, where an attacker attempts to build a long chain of Sybil nodes to reach the seed nodes. As a result, transitivity decay directly addresses PHT’s vulnerability to serial attacks.

Due to its superior performance in mitigating Sybil attacks, we use PHT as the base algorithm for all later experiments. For PHT, we select the serial attack as the most effective one observed in prior tests. For the subsequent experiments, we only report the repeated Sybil attack, as it has more pronounced effects on the network’s reputation system.

\begin{figure}
	\center
	\includegraphics[width=0.65\linewidth]{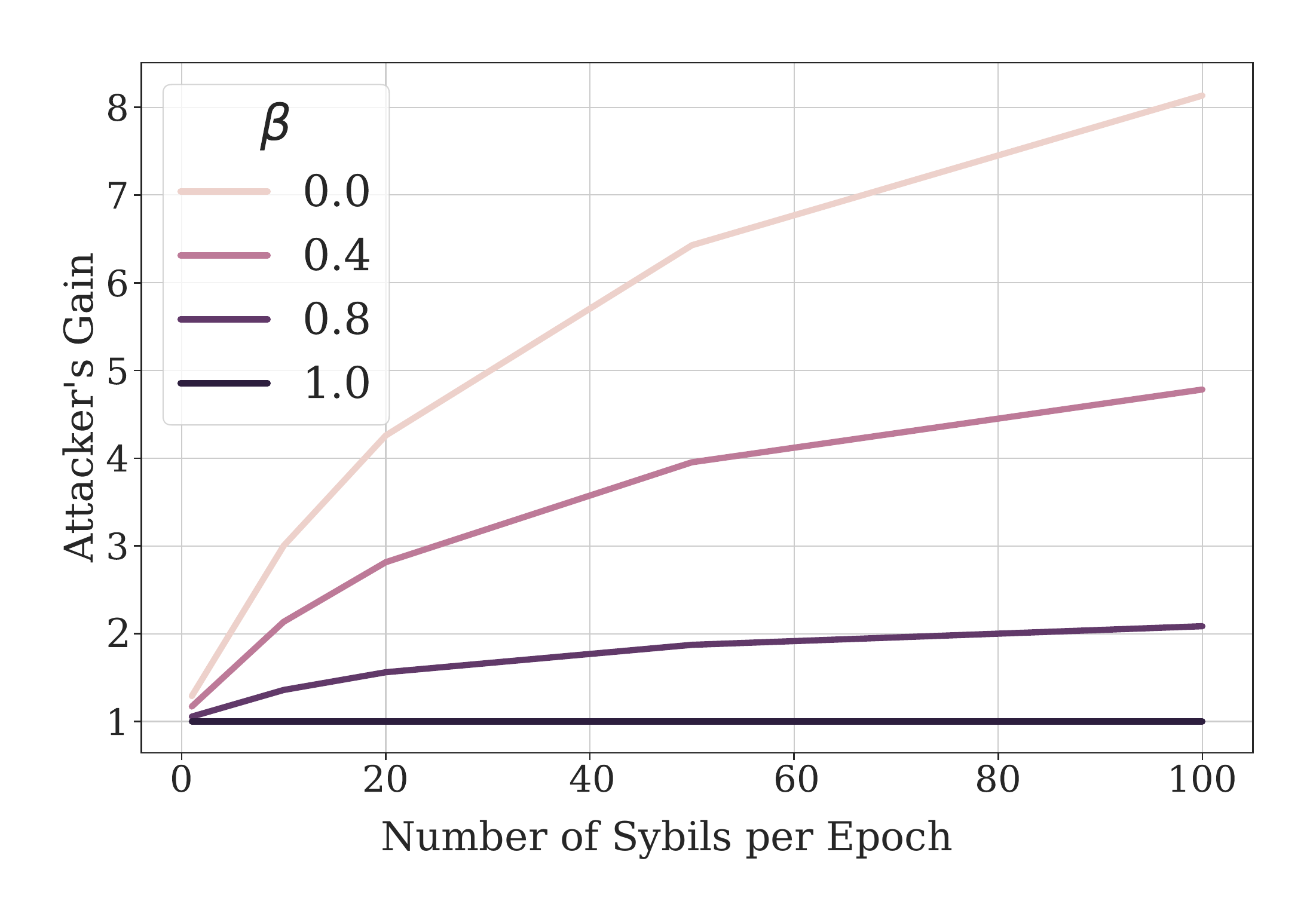}
	\caption{The attacker's gain versus the number of Sybils added per epoch given the connectivity decay $\beta$ with the reputation algorithm PHT.  The connectivity decay is fixed at $\alpha = 0.05$. }
	\label{fig:beta_bounds}
\end{figure}

\subsection{Connectivity Decay}

For the connectivity decay experiments, we set the transitivity decay coefficient to a small value of $\alpha = 0.05$ and the connectivity threshold to $t = 1$. This configuration targets nodes that are connected to the graph exclusively through a single intermediary node, a typical characteristic of Sybil nodes. This decay is also spatial in its nature, like the transitivity decay. 

Figure~\ref{fig:beta_bounds} illustrates the impact of connectivity decay on the attacker’s gain. When there is no connectivity decay ($\beta = 0$), the attacker’s gain increases sub-linearly with the number of Sybil nodes. However, as the connectivity decay coefficient $\beta$ increases, the attacker’s gain decreases significantly. Notably, with $\beta = 1.0$, the attacker’s gain is completely eliminated because the reputation scores of Sybil nodes connected solely through a single intermediary drop to zero.

\begin{figure}
	\center
	\includegraphics[width=0.65\linewidth]{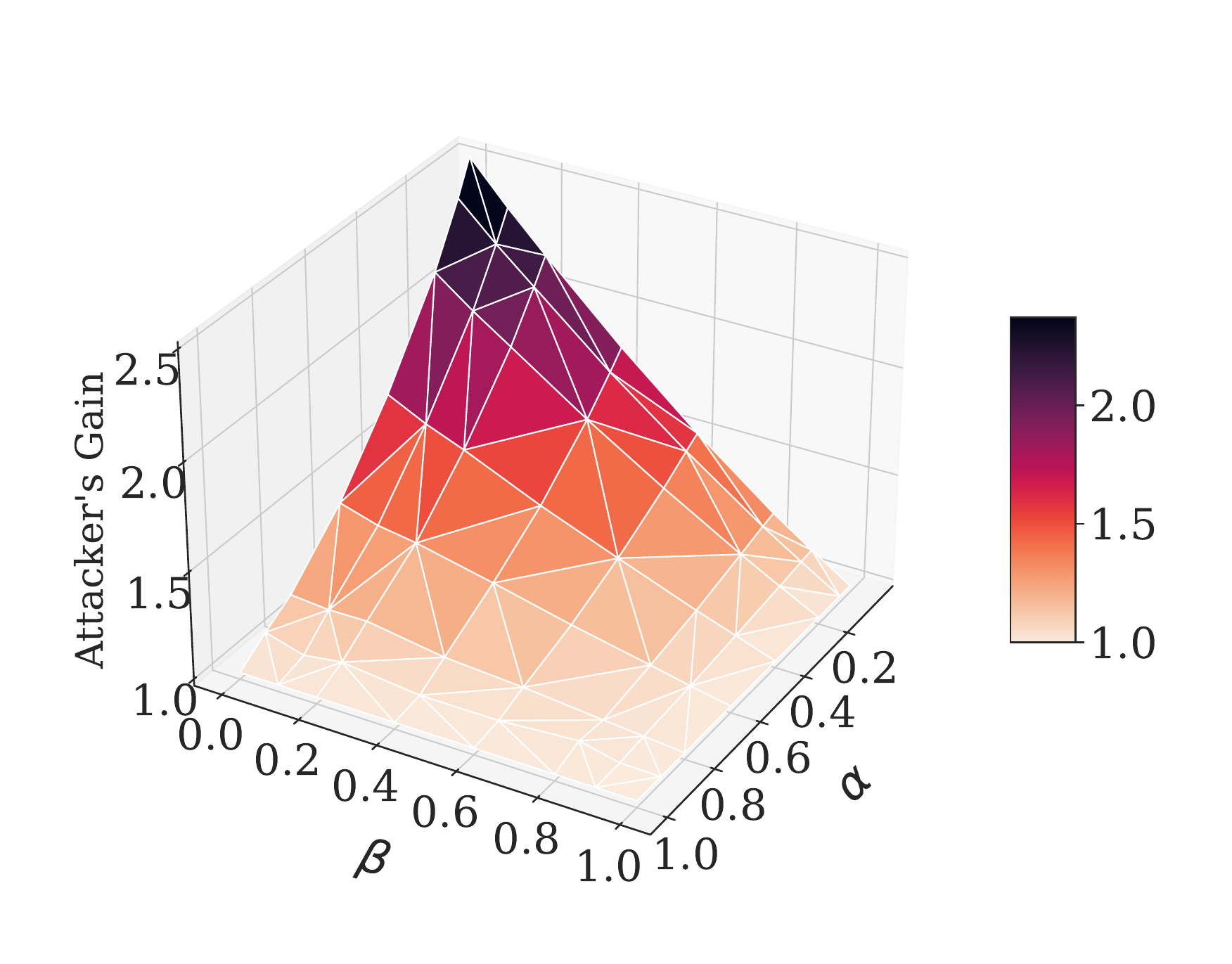}
	\caption{The joint effect of the transitivity decay ($\alpha$) and connectivity decay ($\beta$)  mechanisms on the attacker's gain given $50$ Sybil nodes added per epoch with the reputation algorithm PHT.}
	\label{fig:cum_bounds}
\end{figure}

In Figure~\ref{fig:cum_bounds}, we show that combining connectivity and transitivity decay mechanisms results in greater Sybil tolerance compared to applying each mechanism individually. Specifically, when both the transitivity and connectivity decay coefficients are set to $0.3$, the attacker’s gain remains limited to $1.5$.

Our experiments demonstrate that \meritrank{} effectively mitigates the attacker’s gain through spatial decay mechanisms, even against a significant repeated Sybil attack. Transitivity decay penalizes nodes that are too distant from the seed node, while connectivity decay penalizes nodes that are sparsely connected. Consequently, transitivity decay reduce s the effectiveness of serial attacks, while connectivity decay addresses parallel and cycle attacks. By combining both mechanisms, \meritrank{} achieves Sybil tolerance as defined in Definition~\ref{def:tolerance}, with a threshold of $1 \leq c \leq 2$.

\subsection{Epoch  Decay}

\begin{figure}
	\center
	\includegraphics[width=0.65\linewidth]{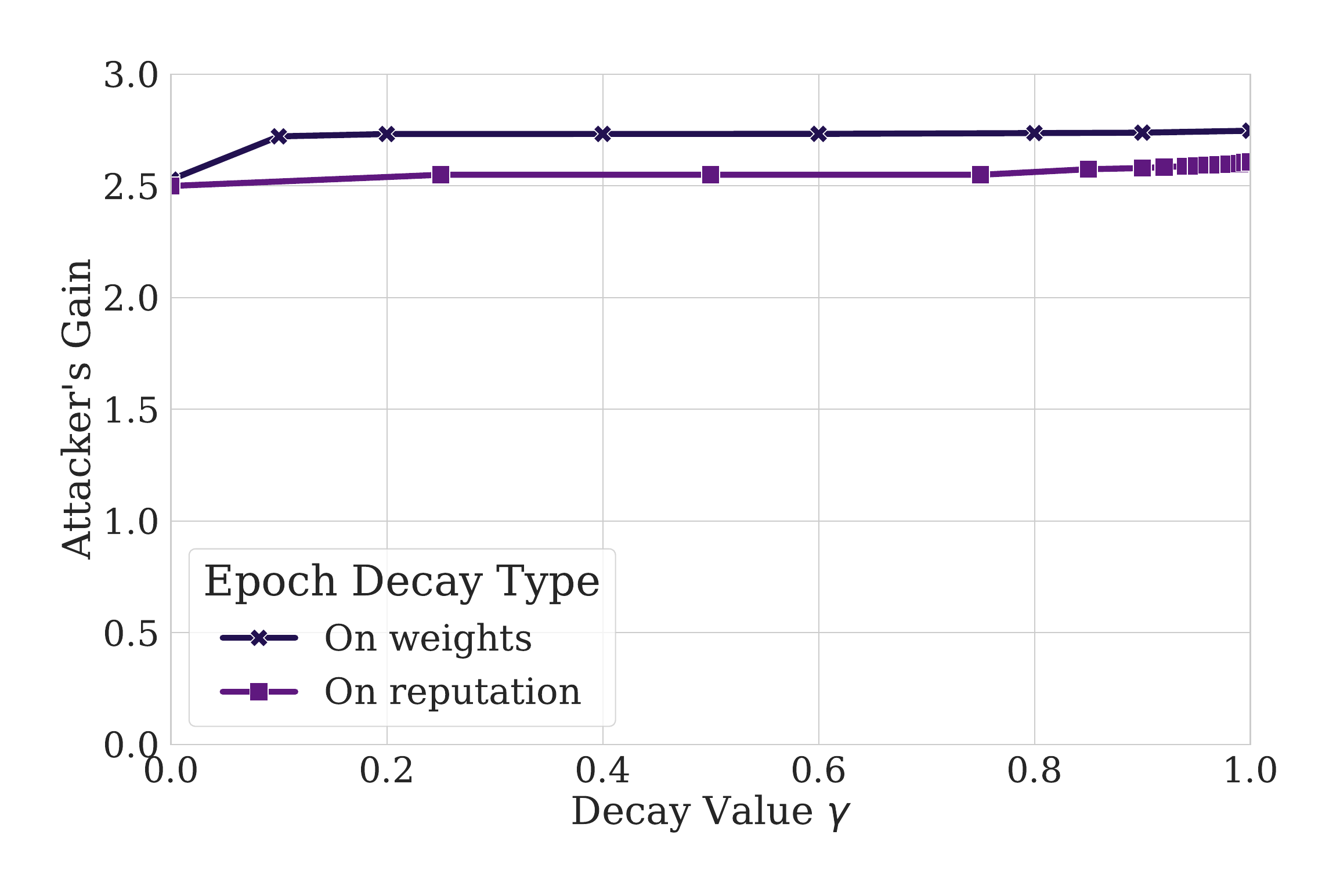}
	\caption{The attacker's gain versus epoch decay ($\gamma$) with $50$ Sybil nodes added per epoch with reputation algorithm PHT.}
	\label{fig:gamma_bounds}
\end{figure}

We conduct experiments to evaluate the effectiveness of epoch decay in mitigating Sybil attacks by applying it to both reputation values and edge weights. The experiments simulate a repeated Sybil attack scenario with $50$ Sybil nodes introduced per epoch, while the transitivity decay coefficient is fixed at $\alpha = 0.05$. The results, presented in Figure~\ref{fig:gamma_bounds}, show the attacker’s gain across different epoch decay values. The findings indicate that epoch decay has minimal impact on the attacker’s gain, regardless of whether it is applied to reputation values or graph weights. In both cases, the attacker’s gain remains consistently high, with only minor fluctuations as the decay value increases.

A notable insight from our experiments is that epoch decay fails to address repeated Sybil attacks effectively. The underlying intuition behind epoch decay is to limit the cumulative influence of nodes over time by penalizing their contribution across epochs. This approach is intended to counter specific attack strategies where an attacker invests in creating a few attack edges and repeatedly exploits them by generating additional Sybil regions. However, our results demonstrate that this mechanism can be easily circumvented by maintaining at least one attack edge and introducing new Sybil nodes each epoch. As a result, the cumulative gain from the attack remains unaffected, rendering epoch decay ineffective in mitigating repeated Sybil attacks.

\subsection{Effect of Decays on Informativeness}~\label{sec:informativeness}

\begin{figure}
	\center
	\includegraphics[width=0.65\linewidth]{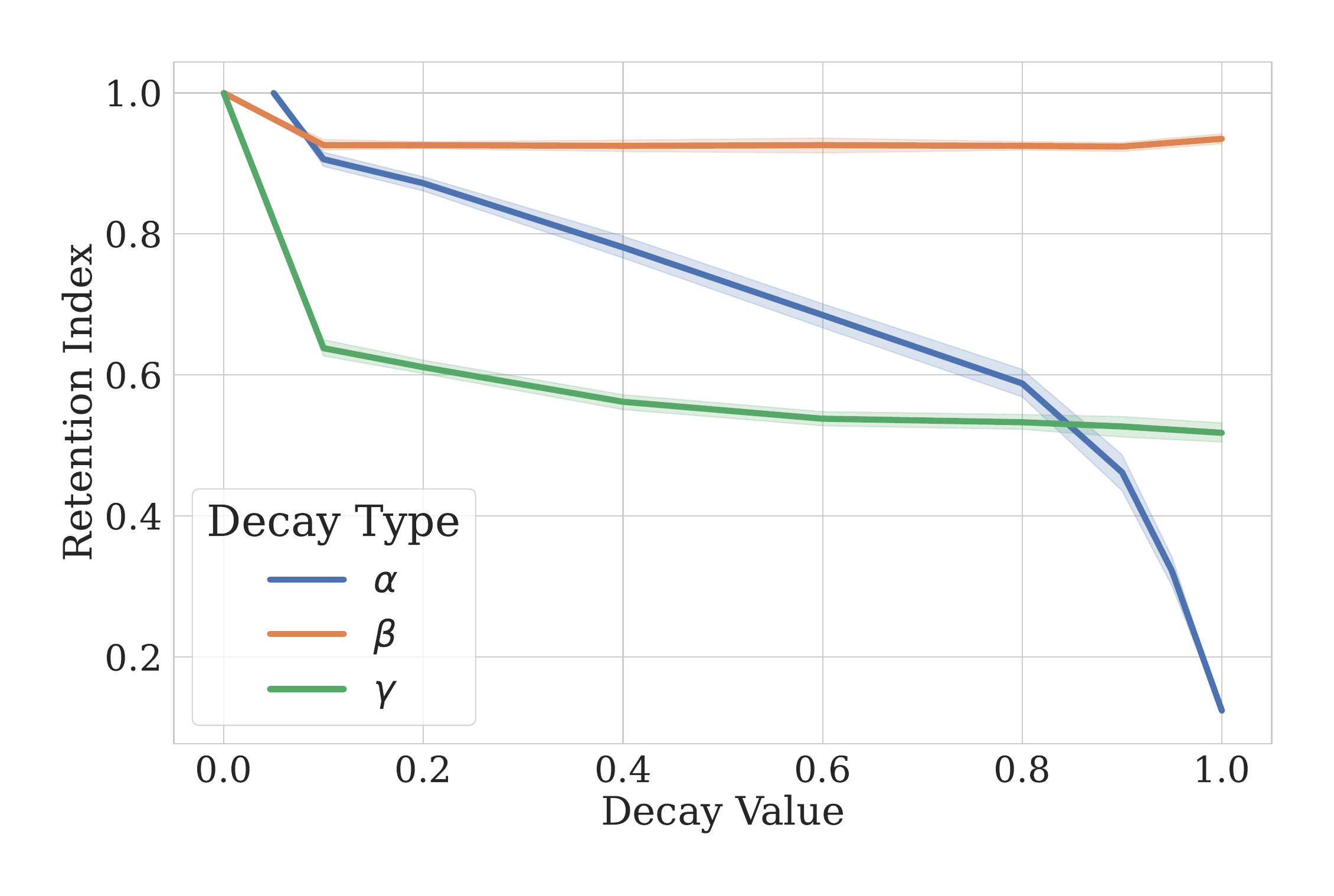}
	\caption{The Retention Index for the top 100 nodes given a reputation algorithm PHT versus the decay values for the decay mechanisms:  transitivity ($\alpha$), connectivity ($\beta$) and epoch ($\gamma$).}
	\label{fig:retention}
\end{figure}

In the previous experiments, we demonstrated how decay mechanisms can effectively limit Sybil attacks. However, these mechanisms are applied universally, even in the absence of Sybil nodes, which can unintentionally degrade the reputation scores of honest participants. Specifically, transitivity decay favors participants close to highly reputable nodes, connectivity decay prioritizes well-connected nodes, and epoch decay penalizes participants who are not consistently active. To quantify this unintended impact, we introduce a metric called \emph{informativeness}, which evaluates how well the original reputation scores of honest participants are preserved after applying decay. To measure informativeness, we execute our simulation applying decay mechanisms without executing any Sybil attacks, reporting the informativeness at the end of epoch 153. We fix $t = 1$ for connectivity decay.

We use two complementary metrics to measure informativeness: the Retention Index for the top N nodes and the Mean Proportional Deviation (MPD). The Retention Index measures how many of the original top $N$ nodes remain in the top $N$ ranking after applying decay. This is particularly useful in scenarios where maintaining the composition of top contributors is important, for example, an allocation score that distributes all rewards to the top 100 most reputable nodes. The MPD, on the other hand, measures the exact deviations in reputation values. This metric is valuable in situations where rewards are allocated proportionally to reputation scores.

The Retention Index quantifies the preservation as the intersection of set of top $N$ nodes before and after applying decay.  This metric ranges from $0$ (no overlap between the sets) to $1$ (complete overlap). Let $N$ denote the number of top nodes considered, and let $d \in [0,1]$ represent the value of the decay parameter for a given decay mechanism. We define $V_N^d \subseteq V$ as the set of the top $N$ nodes with the highest reputation scores after applying decay, and $V_N^0 \subseteq V$ as the corresponding set before applying decay. The Retention Index, $R(d)$, is defined as:

\begin{equation*}
R(d) = \frac{|V_{N}^{0} \cap V_{N}^d|}{|V_{N}^{0}|}    
\end{equation*}

Figure~\ref{fig:retention} illustrates the Retention Index for the top 100 nodes versus the decay value for all three decay mechanisms. The results indicate that connectivity decay ($\beta$) maintains the highest overlap with the baseline ranking across all decay values. Transitivity decay ($\alpha$) exhibits a steady linear decline in overlap as decay increases, with higher decay values leading to a significant divergence from the baseline ranking. Epoch decay ($\gamma$) shows a significant drop at a decay value of 0.1. With this value 63 \% of the nodes retain their rankings in the top 100. The index continues to decrease with further decay but stabilizes, maintaining approximately 50 \% overlap when $\gamma = 1.0$.

The decline in the Retention Index with increasing decay is a direct result of the mechanisms prioritizing nodes closer to the seed, those with stronger connectivity, or participants with consistent activity. Our results indicate that informativeness largely depends on the connectivity and activity levels of participants. In the case of MakerDAO, the top 100 nodes are densely connected and highly active, with approximately 60\% participating in at least half of the epochs. This suggests that the decay mechanisms have minimal impact on informativeness for the most central and consistently active participants, while peripheral or less active nodes may experience greater penalties.

\begin{figure*}[t]
\centering
\begin{subfigure}{0.47\textwidth}
\centering
\includegraphics[width=1.0\textwidth]{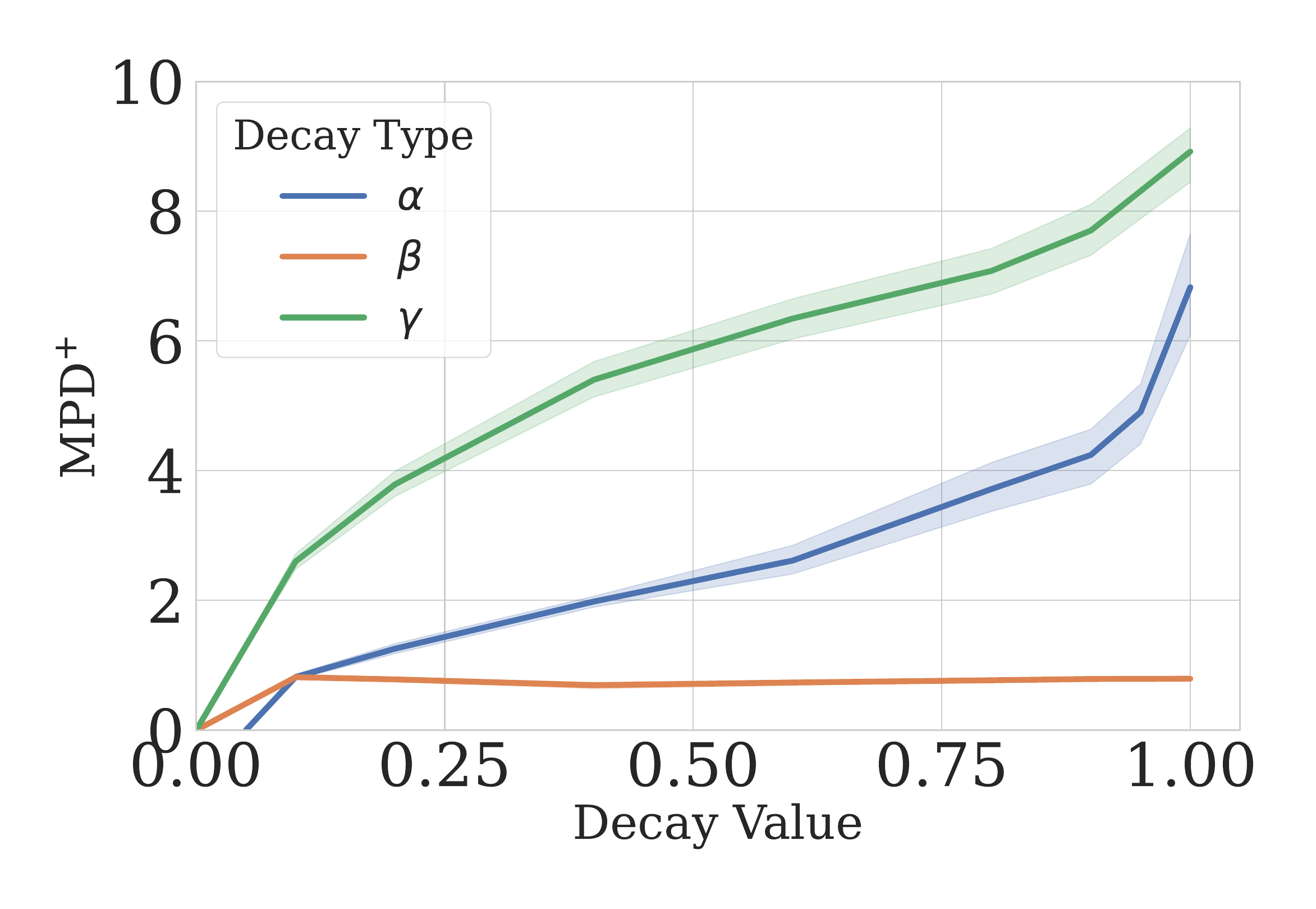}
\end{subfigure} \quad
\begin{subfigure}{0.47\textwidth}
\centering
\includegraphics[width=1.0\textwidth]{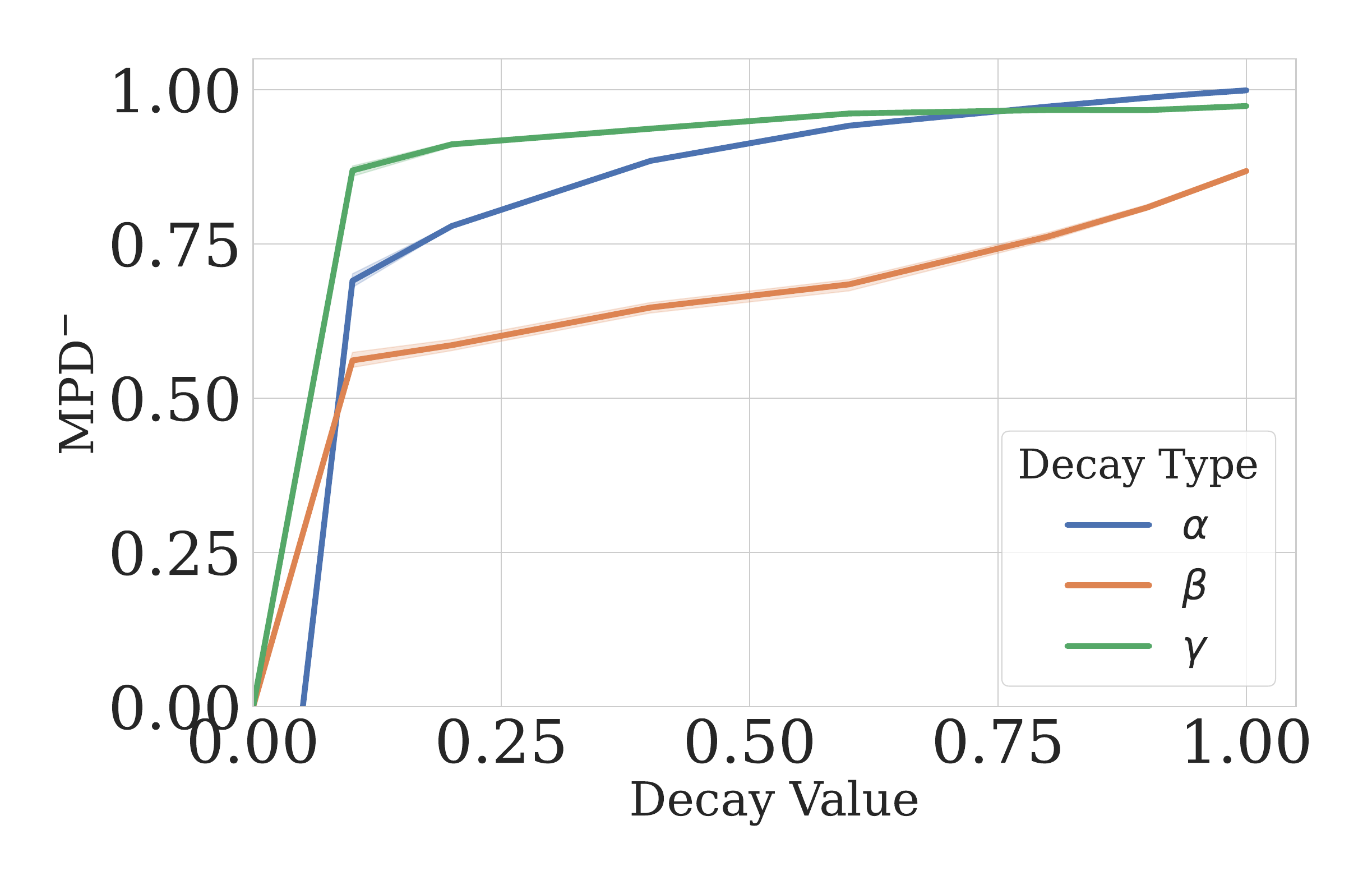}
\end{subfigure}\quad
\caption{The positive and negative mean proportional deviation ($MPD^{+}$ and $MPD^{-}$) versus the decay value for the three decay mechanisms with the reputation algorithm PHT.}
\label{fig:informativeness}
\end{figure*}

Our second metric, MPD, quantifies the deviation of reputation scores from their original values when decay mechanisms are applied. For a given node \( i \in V \), the proportional deviation of the reputation score \( R_i \) under a decay parameter \( d \) is defined as follows:

\begin{equation*}
\delta_{i}(d, k) = \frac{R_{i, decay = d}(G_i, k) - R_{i, decay = 0}(G_i, k)}{R_{i, decay = 0}(G_i, k)}  
\end{equation*}

The MPD is then separated into its positive and negative components. $MPD^{+}(d)$  captures the positive deviations (increases in reputation) and  $MPD^{-}(d)$  captures the absolute values of the negative deviations (decreases in reputation):

\begin{equation*}
MPD^+(d) = \frac{1}{|\{ k \in V \setminus \{i\} \mid \delta_i(d, k) > 0 \}|} \sum_{\substack{k \in V \setminus \{i\} \\ \delta_i(d, k) > 0}} \delta_i(d, k),
\end{equation*}

\begin{equation*}
MPD^-(d) = \frac{1}{|\{ k \in V \setminus \{i\} \mid \delta_i(d, k) < 0 \}|} \sum_{\substack{k \in V \setminus \{i\} \\ \delta_i(d, k) < 0}} |\delta_i(d, k)|.  
\end{equation*}

Figure~\ref{fig:informativeness} illustrates the behavior of $MPD^{+}(d)$ and $MPD^{-}(d)$ under the three decay mechanisms, averaged across 10 independent runs. The results demonstrate a clear divergence in how the different decay mechanisms influence the informativeness of the reputation scores.
 
For positive deviations ($MPD^{+}$), both transitivity decay and epoch decay exhibit a consistent and pronounced increase as decay values rise. In the case of transitivity decay, as the decay value increases, nodes closer to the seed node accrue disproportionately high reputation scores relative to their baseline. At the maximum decay value ($\alpha = 1.0$), direct neighbors of the seed node experience an average increase in their reputation scores by a factor of $7$. A similar trend is observed for epoch decay, resulting in high deviations even with smaller decay values.

In contrast, connectivity decay shows a relatively stable and flat trend after an initial increase. This stability arises because connectivity decay selectively penalizes nodes with low local connectivity (i.e., nodes that are reachable through only a single path given $t$ is $1$). In the MakerDAO graph 23 \% of the nodes have 0 or 1 incoming connections, which would be target for the penalty.

For negative deviations ($MPD^{-}$), all three decay mechanisms exhibit a rapid initial increase, followed by slower, steady growth as the decay value rises. At $\alpha = 1.0$, transitivity decay leads to the complete nullification of all reputation scores, except for those of the direct neighbors of the seed node, leading to $\delta_i(1.0, k) = -1$. A similar effect is observed with epoch decay, where older contributions are systematically diminished, resulting in a substantial degradation of all reputation scores. In contrast, the increase in $MPD^{-}$ for connectivity decay is less pronounced, this is because it does not uniformly reduce reputation scores across the graph. Specifically, $\delta_i = -1$ for the nodes that are sparsely connected, and $\delta_i = 0$ for the others.

Our experiments reveal that connectivity decay preserves informativeness better than transitivity and epoch decay for both the Retention Index and MPD metrics. This outcome is expected, as the connectivity decay affects a smaller subset of nodes within the graph.

\section{Conclusion}

The advances in complex blockchain applications have exposed the limitations of naive tokenomics, which often rely on simple models of monetary incentives for token holders. Merit-based reputation schemes that reward active contributors represent a promising direction for the development of novel blockchain applications such as DAOs. However, the application of reputation in decentralized environments is constrained by the \textit{reputation trilemma}, which posits that a system cannot simultaneously be generalizable, trustless, and Sybil-resistant. We argue that feedback aggregation mechanisms offer the best approach to overcome this challenge as they maintain the trustless property essential to decentralization and are generalizable across various contexts. However, these mechanisms are inherently vulnerable to Sybil attacks, where malicious actors can inflate their reputation scores by creating multiple fake identities. To address this issue, we propose \meritrank{}, Sybil-tolerant reputations based on feedback aggregation with three decay mechanisms— transitivity decay, connectivity decay, and epoch decay. 

Using a dataset of MakerDAO participant interactions, we experimentally demonstrate that \meritrank{} effectively limits the gains of attackers employing Sybil strategies. Furthermore, we investigate informativeness, i.e. the ability to accurately preserve the reputation of honest participants. Our experiments indicate that a combination of transitivity decay and connectivity decay achieves a desirable level of Sybil tolerance while preserving higher informativeness. In contrast, our findings reveal that epoch decay does not enhance Sybil tolerance and can even be counterproductive, as attackers can exploit it by continuously introducing new Sybil identities.

In conclusion, \meritrank{} offers a practical and effective solution for addressing the reputation trilemma. It allows for the application of different decay parameter choices to balance the trade-offs between informativeness and Sybil tolerance. Future work could explore alternative heuristics to optimize this balance and further investigate the counterproductive effects of certain mechanisms, such as epoch decay, to mitigate unintended vulnerabilities in reputation systems.

\bibliographystyle{IEEEtran}
\bibliography{references}
\end{document}